\documentclass{article}

\usepackage{arxiv}
\usepackage{algorithm}
\usepackage{algorithmicx}
\usepackage[spaceRequire=true,indLines=false]{algpseudocodex}
\usepackage{algpseudocode}

\usepackage{color}
\usepackage{amsmath}
\usepackage{amssymb}
\usepackage{pgfplots}
\usepackage{cite}
\usepackage{bm}

\floatname{algorithm}{Algorithm}

\usepackage{color}
\usepackage{setspace}
\usepackage{multirow}
\usepackage{booktabs, makecell}

\usepackage{graphicx} 
\usepackage{float}
\usepackage{subfigure}

\begin{document}
%

\title{Poisson Flow Joint Model for
Multiphase contrast-enhanced CT}
%
%
%

\author{
Rongjun Ge \\ 
Biomedical Imaging Center\\ 
Rensselaer Polytechnic Institute\\
Troy, NY, USA\\
ger@rpi.edu; rongjun\_ge@seu.edu.cn
\And
Ge Wang \\
Biomedical Imaging Center\\ 
Rensselaer Polytechnic Institute\\
Troy, NY, USA\\
wangg6@rpi.edu
}

\maketitle

\begin{abstract}
In clinical practice, multiphase contrast-enhanced CT
(MCCT) is important for physiological and pathological imaging with contrast injection, which undergoes non-contrast, venous, and delayed phases. Inevitably, the accumulated radiation dose to a patient is higher for multiphase scans than for a plain CT scan. Low-dose CECT is thus highly desirable, but it often leads to suboptimal image quality due to reduced radiation dose. Recently, a generalized Poisson flow generative model (PFGM++) was proposed to unify the diffusion model and the Poisson flow generative models (PFGM), and outperform either of them with an optimized dimensionality of the augmentation data space, holding a significant promise for generic or conditional image generation.
In this paper, we propose a Poisson flow joint model (PFJM) for low-dose MCCT to suppress image noise and preserve clinical features.
Our model is built on the PFGM++ architecture to transform the multiphase imaging problem into learning the joint distribution of routine-dose MCCT images by optimizing a task-specific generation path with respect to the dimensionality $D$ of the augmented data space. 
Then, our PFJM model takes the joint low-dose MCCT images as the condition and robustly drives the generative trajectory towards the solution in the routine-dose MCCT domain.  
Extensive experiments demonstrate that our model is favorably compared with competing models, with MAE of 8.99 HU, SSIM of 98.75\% and PSNR of 48.24db, as averaged over all the phases.  
\end{abstract}


%

\section{Introduction}
\begin{figure}
    \centering
    \resizebox{0.6\textwidth}{!}
    {\includegraphics{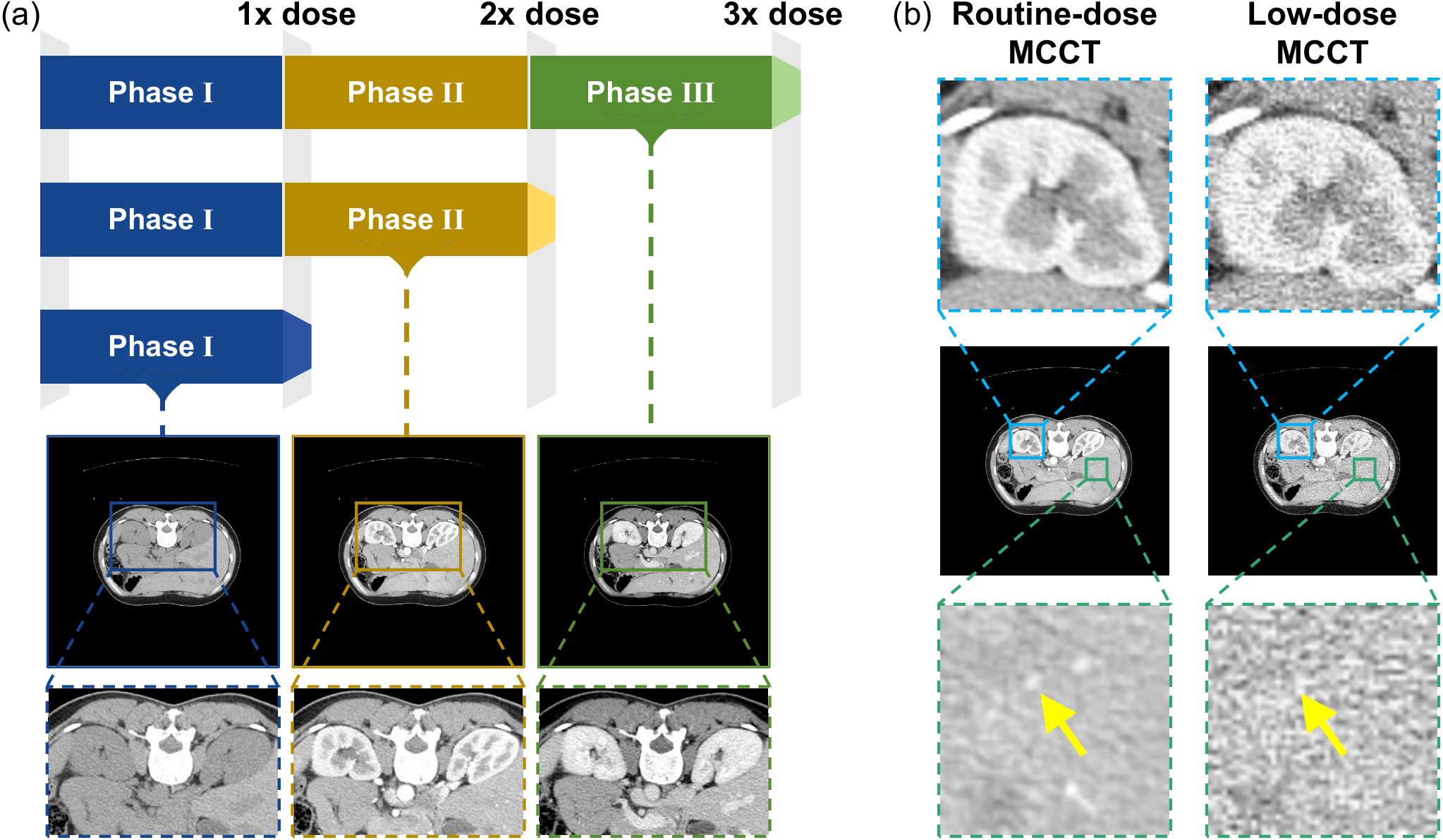}}
    \caption{Image quality challenge with low-dose MCCT.
 (a) The low-dose MCCT workflow, and (b) exemplary images and zoom-in views, showing image degradation at a low radiation dose level.}
    \label{Dose}
\end{figure}

%
%
%
%
Multiphase contrast-enhanced CT (MCCT) plays a significant role in diagnostic imaging. It reveals contrast dynamics and extracts pathologial information \cite{Meng2020}. This contrast enhancement process generally consists of non-contrast, arterial, and venous phases \cite{Smithuis}, which are referred to as phases I, II, III respectively. However, multiple CT scans during these phases lead to accumulation of x-ray radiation dose into a patient \cite{Brenner,Rastogi}, as shown in Fig.~\ref{Dose}(a).  
By the “as low as reasonably achievable” (ALARA) principle \cite{Prasad}, it is necessary to minimize the 
radiation dose and associated risk for the patient in his/her MCCT examination. 
However, low-dose MCCT is subject to poor signal-to-noise ratio and strong image noise that may often compromise  quality of reconstructed images and clarity of clinical features of interest, as shown in Fig.~\ref{Dose}(b). This physical limitation challenges image interpretation, even for experienced radiologists. 
Therefore, there is an important and immediate need for high-quality image reconstruction and post-processing in the scenario of low-dose MCCT.

\begin{figure*}[t]
    \centering
    \resizebox{0.85\textwidth}{!}
    {\includegraphics{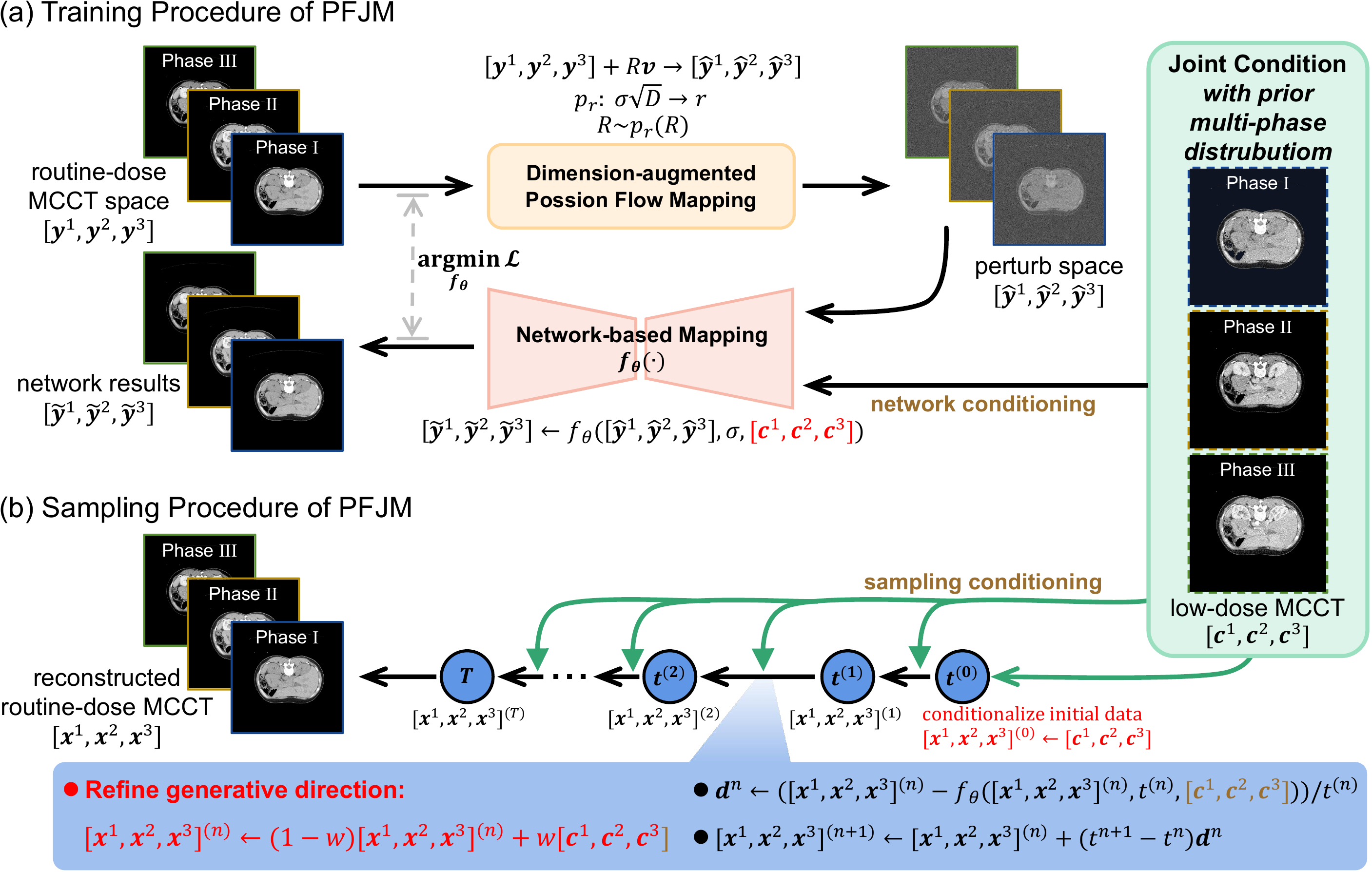}}
        \caption{Overview of the proposed PFJM for low-dose MCCT imaging through the three phases. It focuses on learning the joint distribution of routine-dose (RD) MCCT. With the adjustable augmented dimension $D$, PFJM adaptively optimizes a task-specific generation path for the joint distribution of RD MCCT. The joint condition consistent with prior multi-phase distribution is further leveraged for both network learning and generative sampling. In the learning process, the joint condition of low-dose MCCT images facilitates suppression of hallucinations and blind spots  with the correlated anatomy features. In the sampling process, the joint condition refines the generative direction at each sampling step for superior image reconstruction.
        }
    \label{Overview}
\end{figure*}

During the past decade, deep learning methods have shown success in low-dose CT reconstruction. According to the training mechanism, these methods can be classified into 1) end-to-end supervision and 2) fine-tuning or generative AI-based.
In the first category, 
low-dose CT denoising relies on pixel-to-pixel supervision to learn each pixel value in routine-dose CT \cite{RED_CNN,Yin2019,Fan2019,Lu2022,Yang2022}.
The results with the least-square loss often show an over-smoothing effect. Based on these results, domain adaption via fine-turning cannot fix the over-smoothing problem.
With a generative AI model, a data distribution can be learned \cite{Yang2018,Bera2021,Kwon2021,Moghari2021,Shan2018,Huang2021,Ge2019,Fu2023} to ensure realistic
denoising results. The generative adversarial network (GAN), as a popular framework for generative AI\cite{GAN}, drives the generator to reconstruct high-quality CT images that 
cheat the discriminator effectively. 
However, the adversarial training usually suffers from unstable learning \cite{Salimans} and mode collapse \cite{Metz}, requiring a proper design of the network architecture and an extra care in the training strategy  \cite{Zhao_empirical,Gao_TMI2023}. 

Recently, physics-inspired generative AI models, especially diffusion \cite{Song2020,Karras2022,DDIM} and Poisson flow generative models (PFGM) \cite{PFGM,PFGMPP}, show impressive performance for unconditional \cite{Ho2020nips,Nichol2021PMLR,Song2021ICLR} and conditional \cite{Zhao2023,Song2021,Saharia2022,Saharia2022PAMI,Zhang2023CVPR,Ho2022} image generation. 
While the diffusion model was inspired by non-equilibrium thermodynamics, PFGM was derived based on electrostatics.
Such generative models are provably capable of learning complex data distributions, allowing stable training, strong mode coverage, high-quality sampling, and flexible conditioning \cite{Kazerouni}. 
This contemporay approach has thus gained much attention in the field of medical imaging \cite{Hein2024,Du2024,GeMICCAI2023,Hein2,Restrepo,Karageorgos,Guan,Liu_ICCV}. 

\begin{figure}[b]
    \centering
    \resizebox{0.5\textwidth}{!}
    {\includegraphics{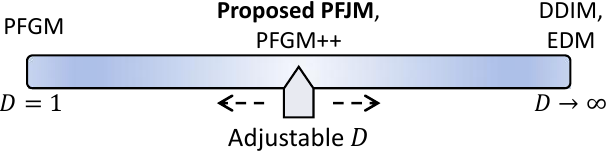}}
    \caption{Based on PFGM++, PFJM flexibly adjusts $D$ to optimize the path for image generation in a task-specific fashion, outperforming in general either PFGM ($D=1$) or diffusion model ($D\rightarrow\infty$).  
           }
    \label{Ajustable}
\end{figure}

The diffusion/PFGM methods take routine-dose CT images as the starting point of the forward procedure to reach a feature-free field \cite{Du2024,GeMICCAI2023,Hein2} and then learn to map a sample of the feature-free field to an image under the routine-dose CT distribution. When a low-dose CT image is used as the condition to control the generated image content, and the outcome will be a denoised image consistent to the target distribution.
By unifying the diffusion and PFGM models, PFGM++ \cite{PFGMPP} enables 
flexible generative paths in a task-specific fashion by adjusting the augmented dimension $D$. 
The key idea is to embed generative trajectories for $N$ dimensional data in an $N+D$ dimensional space while  controlling the degenerative process with a scalar norm of the vector of $D$ augmented variables. 
As proven in \cite{PFGMPP}, PFGM++ becomes PFGM for $D=1$ and the diffusion model for $D\rightarrow\infty$. 
By choosing an appropriate $D$ in PFGM++, a task-specific generation path can be optimized, outperforming either the diffusion or Poisson flow models in general and holding a great potential in medical imaging applications.


In this paper, we presents a  
Poission flow joint model (PFJM) for low-dose MCCT imaging through all the three phases, as shown in Fig.~\ref{Overview}. 
Our proposed model promises great potential to reduce radiation dose to the patient undergoing an MCCT scan without compromising the diagnostic performance. 
Our work is based on PFGM++ and focused on learning the joint distribution of routine-dose MCCT images. 
Like PFGM++, our proposed PFJM model treats the data distribution as a distribution of electric charges in an augmented space of dimensionality $N+D$ where the augmented dimension $D$ can be adjusted to facilitate modeling of the original $N$-dimensional data distribution. As shown in Fig.~\ref{Ajustable}, by adjusting $D$, the generation path becomes adaptive and can be optimized in the context of denoising MCCT images.
Also, to control the manipulation of MCCT images, we employ the multiphase joint condition in both network learning and generative sampling. In the learning process, the joint condition of low-dose MCCT images is injected into the network so that the correlated anatomy features can facilitate suppression of hallucinations and blind spots to speed up the convergence. In the sampling process, the joint condition of MCCT images helps refine the generative direction for each sampling step.

Our main contributions are summarized as follows:
\begin{itemize}
\item 
Our proposed PFJM is the first``PFGM++'-based framework for low-dose MCCT imaging, which flexibly adjusts $D$ to optimize the generative path in a task-specific fashion.  
\item 
Our training strategy utilizes the multi-phase joint condition in both network learning and generative sampling, precisely controling the generative process and denoised image quality, by embedding the prior distribution of correlated anatomic features across multiple phases. 
\item 
The augmented dimensionality $D$ has been systematically optimized in extensive experiments, giving encouraging feasibility results.
\end{itemize}

    \begin{algorithm}[b]
        \caption{Training Procedure of the Proposed PFJM}
        \begin{algorithmic}[1] %
            \Require  
                      Routine dose (RD) MCCT Dataset $(\mathcal{Y}^1$, $\mathcal{Y}^2$, $\mathcal{Y}^3)$; 
                      Low-dose (LD) MCCT Dataset $(\mathcal{C}^1$, $\mathcal{C}^2$, $\mathcal{C}^3)$
            \Ensure Trained Model $f_\theta$  
            \For{Each iteration} 
                \State Sample RD MCCT images  $\left [ \textit{\textbf{y}}^1,\ \textit{\textbf{y}}^2,\ \textit{\textbf{y}}^3 \right ]$ from $\mathcal{Y}^1$, $\mathcal{Y}^2$, and $\mathcal{Y}^3$
                \State Sample standard deviation $\sigma$ from $p(\sigma)$
                \State Sample $r$ from $p_r$: $r \gets \sigma \sqrt{D} $
                \State Sample $R\sim p_{r(R)}$
                \State Sample angle $\textit{\textbf{v}}\gets\frac{ \textit{\textbf{u}}}{\left \| \textit{\textbf{u}} \right \| _2} $, with $\textit{\textbf{u}}\sim \mathcal{N}(\textbf{0},\textbf{I}) $
                \State Perturb data $[\hat{\textit{\textbf{y}}}^1,\hat{\textit{\textbf{y}}}^2,\hat{\textit{\textbf{y}}}^3] \gets [\textit{\textbf{y}}^1,\textit{\textbf{y}}^2,\textit{\textbf{y}}^3]+R_i\textit{\textbf{v}}$
                \State \textcolor{red} {Sample LD MCCT images $\left [ \textit{\textbf{c}}^1,\ \textit{\textbf{c}}^2,\ \textit{\textbf{c}}^3 \right ]$ from $\mathcal{C}^1$, $\mathcal{C}^2$, and $\mathcal{C}^3$}
                \State \textcolor{red} {Build the joint condition $[\textit{\textbf{c}}_i^1,\ \textit{\textbf{c}}_i^2,\ \textit{\textbf{c}}_i^3]$}
                \State Calculate the loss \\ $\mathcal{L}\! \gets\! \lambda (\sigma )\!\left \| f_{\theta}([\hat{\textit{\textbf{y}}}^1,\hat{\textit{\textbf{y}}}^2,\hat{\textit{\textbf{y}}}^3],\sigma,\left [ \textit{\textbf{c}}^1,\textit{\textbf{c}}^2,\textit{\textbf{c}}^3 \right ])\!-\![\textit{\textbf{y}}^1,\textit{\textbf{y}}^2,\textit{\textbf{y}}^3]  \right \|_2^2  $
                \State Update the model $f_\theta$ with the Adam optimizer
                \EndFor           \\  
            \Return $f_\theta$
        \end{algorithmic}
        \label{Training}
    \end{algorithm}

\section{Methodology}
\subsection{Formulation of the PFGM++ Model}
PFGM++\cite{PFGMPP}, a new family of physics-inspired generative models, is inspired by the theory of electrostatics.
It interprets an $N$-dimensional datum $\textit{\textbf{x}}\in \mathbb{R}^N$ as an electric charge in the original $N$-dimensional space augmented with additional $D$ dimensions so that the augmented counterpart of the datum is $\tilde{\textit{\textbf{x}}}=(\textit{\textbf{x}},\textit{\textbf{z}})\in \mathbb{R}^{N+D}$, where $\textit{\textbf{z}}=(z_1,z_2,...,z_D)$ and $D\in \mathbb{Z}^+$.
The training data are located on the $\textit{\textbf{z}}=\textbf{0}$ hyperplane, and the electric field lines from the charges define a bijection between the data distribution and a uniform distribution on the infinitely large hemisphere in the augmented space.
PFGM++ produces samples by evolving points from the hemisphere back to the $\textit{\textbf{z}}=\textbf{0}$ hyperplane. The main structure of PFGM++ is the high-dimensional electric field expressed as
\begin{equation}
\textit{\textbf{E}}(\tilde{\textit{\textbf{x}}})=\frac{1}{S_{N+D-1}(1)}\int \frac{\tilde{\textit{\textbf{x}}}-\tilde{\textit{\textbf{y}}}}{\left\| \tilde{\textit{\textbf{x}}}-\tilde{\textit{\textbf{y}}} \right\|^{N+D}}p(\textit{\textbf{y}})\mathrm{d}\textit{\textbf{y}}  
\label{Eq1}
\end{equation}
where $p(\textit{\textbf{y}})$ represents the ground truth data distribution, $S_{N+D-1}(1)$ is the surface area of a unit ($N+D-1$)-sphere,  $\tilde{\textit{\textbf{y}}}=(\textit{\textbf{y}},\textbf{0})\in \mathbb{R}^{N+D}$ and $\tilde{\textit{\textbf{x}}}=(\textit{\textbf{x}},\textit{\textbf{z}})\in \mathbb{R}^{N+D}$ are the augmented
ground truth and perturbed data, respectively.   

The electric field has rotational symmetry on the $D$-dimensional cylinder $\sum_{i=1}^{D} z_i^2=r^2$ for any $r>0$. Due to the symmetry, it suffices to track the norm of the vector of augmented variables $r=r(\tilde{\textit{\textbf{x}}})=\left\| \textit{\textbf{z}} \right\|_2$, instead of modeling the individual behavior of each $z_i$. Replacing the notation for augmented data with
$\tilde{\textit{\textbf{x}}}=(\textit{\textbf{x}},r)$, the physically meaningful $r$ can be used as the anchor variable in the ODE:
\begin{equation}
\frac{\mathrm{d}\textit{\textbf{x}}}{\mathrm{d}r} = \frac{\textit{\textbf{E}}(\tilde{\textit{\textbf{x}}})_\textit{\textbf{x}}} {E(\tilde{\textit{\textbf{x}}})_r}
\label{Eq2}
\end{equation}
where $E(\tilde{\textit{\textbf{x}}})_r=\frac{1}{S_{N+D-1}(1)}\int \frac{r}{\left\| \tilde{\textit{\textbf{x}}}-\tilde{\textit{\textbf{y}}} \right\|^{N+D}}p(\textit{\textbf{y}})d\textit{\textbf{y}}$ is a scalar.
The aforementioned surjection is thus turned into a bijection between the distribution on the $r=r_{max}$ hyper-cylinder and the ground truth data distribution on $r=0$ (i.e., the $\textit{\textbf{z}}=\textbf{0}$) hyperplane \cite{PFGMPP,PFGM}. 

PFGM++ employs a perturbation based objective, akin to the denoising score matching objective in diffusion models \cite{Vincent2011}, without using large batches to construct the electric field similar to that used in PFGM \cite{PFGM}. 
The objective is given as
\begin{equation}
\mathbb{E}_{r\sim p(r)}\mathbb{E}_{\textit{\textbf{y}}\sim p(\textit{\textbf{y}})}\mathbb{E}_{\textit{\textbf{x}}\sim p_r(\textit{\textbf{x}}|\textit{\textbf{y}})}\left [ \left \| f_\theta (\tilde{\textit{\textbf{x}}})-\frac{\textit{\textbf{x}}-\textit{\textbf{y}}}{r/\sqrt{D} }  \right \|_2^2   \right ].
\label{Eq3}
\end{equation}

    \begin{algorithm}[h]
        \setstretch{1.1}
        \caption{Sampling Procedure of the Proposed PFJM}
        \begin{algorithmic}[1] %
            \Require  
                      LD MCCT images $\left [ \textit{\textbf{c}}^1,\ \textit{\textbf{c}}^2,\ \textit{\textbf{c}}^3 \right ]$;                      
                      Trained Model $f_\theta$;
                      Sampling step $T$
            \Ensure Reconstructed RD MCCT images  $[\textit{\textbf{x}}^1, \textit{\textbf{x}}^2, \textit{\textbf{x}}^3]$  
            \State \textcolor{red} {Build the joint condition $ [\textit{\textbf{c}}^1,\ \textit{\textbf{c}}^2,\ \textit{\textbf{c}}^3]$}
            \State \textcolor{red} {Obtain the initial data $[\textit{\textbf{x}}^1,\ \textit{\textbf{x}}^2,\ \textit{\textbf{x}}^3]^{(0)} \gets [\textit{\textbf{c}}^1,\ \textit{\textbf{c}}^2,\ \textit{\textbf{c}}^3]$}
            \For {$n=0,...,T-1$} 
                \State \textcolor{red} {Refine the generative direction \\
                $[\textit{\textbf{x}}^1,\textit{\textbf{x}}^2,\textit{\textbf{x}}^3]^{(n)}=(1-w)[\textit{\textbf{x}}^1, \textit{\textbf{x}}^2, \textit{\textbf{x}}^3]^{(n)}+w[\textit{\textbf{c}}^1, \textit{\textbf{c}}^2, \textit{\textbf{c}}^3]$}
                \State $\textit{\textbf{d}}^{(n)}  \gets  [\textit{\textbf{x}}^1, \textit{\textbf{x}}^2, \textit{\textbf{x}}^3]^{(n)}-f_\theta([\textit{\textbf{x}}^1, \textit{\textbf{x}}^2, \textit{\textbf{x}}^3]^{(n)},t^{(n)},[\textit{\textbf{c}}^1,\textit{\textbf{c}}^2, $\\$\textit{\textbf{c}}^3]))/t^{(n)}$
                \State $[\textit{\textbf{x}}^1,\textit{\textbf{x}}^2,\textit{\textbf{x}}^3]^{(n+1)}  \gets  [\textit{\textbf{x}}^1,\textit{\textbf{x}}^2,\textit{\textbf{x}}^3]^{(n)} + (t^{(n+1)}-t^{(n)})\textit{\textbf{d}}^{(n)} $
                \If {$t^{(n+1)}>0$}
                    \State ${\textit{\textbf{d}}}'^{(n)}  \gets ([\textit{\textbf{x}}^1, \textit{\textbf{x}}^2, \textit{\textbf{x}}^3]^{(n+1)}-f_{\theta}([\textit{\textbf{x}}^1,\textit{\textbf{x}}^2, \textit{\textbf{x}}^3]^{(n+1)},$\\$t^{(n+1)}, [\textit{\textbf{c}}^1, \textit{\textbf{c}}^2, \textit{\textbf{c}}^3]))/t^{(n+1)} $
                    \State $[\textit{\textbf{x}}^1,\ \textit{\textbf{x}}^2,\ \textit{\textbf{x}}^3]^{(n+1)} \gets[\textit{\textbf{x}}^1,\ \textit{\textbf{x}}^2,\ \textit{\textbf{x}}^3]^{(n)}+(t^{(n+1)}-t^{(n)})( \frac{1}{2}\textit{\textbf{d}}^{(n)}+\frac{1}{2}{\textit{\textbf{d}}}'^{(n)})$
                \EndIf
            \EndFor     \\
            $[\textit{\textbf{x}}^1, \textit{\textbf{x}}^2, \textit{\textbf{x}}^3] \gets [\textit{\textbf{x}}^1, \textit{\textbf{x}}^2, \textit{\textbf{x}}^3]^{(T)}$  \\
        \Return $[\textit{\textbf{x}}^1, \textit{\textbf{x}}^2, \textit{\textbf{x}}^3]$
        \end{algorithmic}
         \label{Sampling}
    \end{algorithm}

\subsection{Design of the Poisson Flow Joint Model}
Our PFJM model is designed with the “PFGM++” architecture to perceive the joint distribution of routine-dose MCCT images, and conditioned on multi-phase information to refine the generative direction step-wise, as shown in Fig.~\ref{Overview}.
By adjusting the
variable $D$ to embed generative paths in a task-specific fashion, our model can achieve a highly-adaptive reconstruction path for optimal low-dose MCCT imaging. 
The multi-phase condition is implemented utilizing the joint prior distribution among MCCT images with key features naturally correlated across different phases so that the reverse procedure towards the routine MCCT data space can be explicitly guided.

Specifically, for low-dose MCCT imaging, we enable learning of the joint distribution among different phases in our PFJM model. 
Among different phases, there naturally exists correlation among key image features. The joint distribution mapping promotes such inter-phase synergy to facilitate low-dose MCCT denoising. 
In the forward procedure, the routine-dose joint distribution $[ \textit{\textbf{y}}^1, \textit{\textbf{y}}^2, \textit{\textbf{y}}^3 ]\in\mathbb{R}^{L \times W\times 3}$ is formed through  dimension-augmented Possion flow mapping of PFGM++ perturbation \cite{Karras2022,PFGMPP} in the form of $R_iv_i\sim U_\psi (\psi )p_r(R)$.
In the reverse procedure, to explicitly steer the trajectory toward the posterior joint distribution of routine-dose MCCT images, the prior distribution of the corresponding low-dose MCCT images $ [ \textit{\textbf{c}}^1, \textit{\textbf{c}}^2, \textit{\textbf{c}}^3  ]\in\mathbb{R}^{L \times W\times 3}$ is pertinently incorporated, as Fig.~\ref{Overview}(a) shows. The low-dose multiphase distribution $[ \textit{\textbf{c}}^1, \textit{\textbf{c}}^2, \textit{\textbf{c}}^3  ]$, is used as the joint condition to guide the network-based mapping to the routine-dose MCCT image space so that image hallucinations and blind spots can be suppressed. The training procedure of PFJM is described as Algorithm.~\ref{Training}, with the red text highlighting our updates to the generic PFGM++ model.

In the reverse procedure for generative sampling,
the join condition $[ \textit{\textbf{c}}^1, \textit{\textbf{c}}^2, \textit{\textbf{c}}^3  ]$ from low-dose MCCT images initializes the sampling process and refines intermediate results at each sampling step, as Fig.~\ref{Overview}(b) shows. In learning the mapping relationship between different distributions, global differences like image style usually attract more attention from the model due to its high perceptual salience, leading to losing or distoring anatomical details.  
The joint condition $[ \textit{\textbf{c}}^1, \textit{\textbf{c}}^2, \textit{\textbf{c}}^3 ]$ contains original anatomical information. Thus, at each sampling step for denoising, it is desirable to adjust intermediate images $[\textit{\textbf{x}}^1,\textit{\textbf{x}}^2,\textit{\textbf{x}}^3]^{(n)}$ through $(1-w)[\textit{\textbf{x}}^1, \textit{\textbf{x}}^2, \textit{\textbf{x}}^3]^{(n)}+w[\textit{\textbf{c}}^1, \textit{\textbf{c}}^2, \textit{\textbf{c}}^3]$ to 
refine generation direction with imaging structure and details redeemed. The sampling procedure of PFJM is described as Algorithm.~\ref{Sampling}, with the red text highlighting the updates to PFGM++.

\begin{table*}[t]
\centering
\caption{Quantitative analysis of our PFJM model compared with the existing methods under the various settings. MAE, SSIM and PSNR are calculated for each phase.}
\begin{tabular}{llllllllllc}
\bottomrule[1.5pt]
\multicolumn{1}{c|}{\multirow{2}{*}{}}     & \multicolumn{3}{c|}{Phase I}                                                                                                                                                                                                & \multicolumn{3}{c|}{Phase II}                                                                                                                                                                                               & \multicolumn{3}{l|}{Phase III}                                                                                                                                                                                              & Step                \\ \cline{2-10}
\multicolumn{1}{c|}{}                      & \multicolumn{1}{c}{\begin{tabular}[c]{@{}c@{}}MAE\\ (HU)\end{tabular}} & \multicolumn{1}{c}{\begin{tabular}[c]{@{}c@{}}SSIM\\ (\%)\end{tabular}} & \multicolumn{1}{c|}{\begin{tabular}[c]{@{}c@{}}PSNR\\ (db)\end{tabular}} & \multicolumn{1}{c}{\begin{tabular}[c]{@{}c@{}}MAE\\ (HU)\end{tabular}} & \multicolumn{1}{c}{\begin{tabular}[c]{@{}c@{}}SSIM\\ (\%)\end{tabular}} & \multicolumn{1}{c|}{\begin{tabular}[c]{@{}c@{}}PSNR\\ (db)\end{tabular}} & \multicolumn{1}{c}{\begin{tabular}[c]{@{}c@{}}MAE\\ (HU)\end{tabular}} & \multicolumn{1}{c}{\begin{tabular}[c]{@{}c@{}}SSIM\\ (\%)\end{tabular}} & \multicolumn{1}{c|}{\begin{tabular}[c]{@{}c@{}}PSNR\\ (db)\end{tabular}} &                     \\ \hline
\multicolumn{1}{l|}{PFGM}                  & \multicolumn{1}{c}{12.55}                                              & \multicolumn{1}{c}{97.57}                                               & \multicolumn{1}{c|}{45.51}                                               & \multicolumn{1}{c}{11.00}                                              & \multicolumn{1}{c}{98.17}                                               & \multicolumn{1}{c|}{46.15}                                               & \multicolumn{1}{c}{12.83}                                              & \multicolumn{1}{c}{97.44}                                               & \multicolumn{1}{c|}{45.32}                                               & 30                  \\ \hline
\multicolumn{1}{l|}{\multirow{2}{*}{DDIM}} & 12.08                                                                  & 97.60                                                                   & \multicolumn{1}{l|}{45.97}                                               & 13.34                                                                  & 96.89                                                                   & \multicolumn{1}{l|}{45.51}                                               & 13.13                                                                  & 97.38                                                                   & \multicolumn{1}{l|}{45.45}                                               & 10                  \\
\multicolumn{1}{l|}{}                      & 10.41                                                                  & 98.13                                                                   & \multicolumn{1}{l|}{46.80}                                               & 10.21                                                                  & 98.28                                                                   & \multicolumn{1}{l|}{47.17}                                               & 10.03                                                                  & 98.18                                                                   & \multicolumn{1}{l|}{46.99}                                               & 20                  \\ \hline
\multicolumn{1}{l|}{\multirow{2}{*}{EDM}}  & 15.54                                                                  & 96.62                                                                   & \multicolumn{1}{l|}{43.78}                                               & 15.12                                                                  & 96.91                                                                   & \multicolumn{1}{l|}{43.84}                                               & 15.08                                                                  & 96.87                                                                   & \multicolumn{1}{l|}{43.86}                                               & 10                  \\
\multicolumn{1}{l|}{}                      & 12.56                                                                  & 97.59                                                                   & \multicolumn{1}{l|}{45.55}                                               & 11.99                                                                  & 97.93                                                                   & \multicolumn{1}{l|}{45.67}                                               & 11.97                                                                  & 97.89                                                                   & \multicolumn{1}{r|}{45.68}                                               & 20                  \\ \hline
\hline
\multicolumn{11}{c}{PFGM++}                                                                                                                                                                                                                                                                                                                                                                                                                                                                                                                                                                                                                                                                                                                                \\ \hline
\multicolumn{1}{l|}{D=2}                   & 60.79                                                                  & 69.59                                                                   & \multicolumn{1}{l|}{35.33}                                               & 59.98                                                                  & 70.26                                                                   & \multicolumn{1}{l|}{35.36}                                               & 60.50                                                                  & 70.68                                                                   & \multicolumn{1}{l|}{35.35}                                               & \multirow{8}{*}{10} \\
\multicolumn{1}{l|}{D=8}                   & 15.49                                                                  & 96.44                                                                   & \multicolumn{1}{l|}{43.82}                                               & 15.36                                                                  & 96.49                                                                   & \multicolumn{1}{l|}{43.76}                                               & 15.77                                                                  & 96.18                                                                   & \multicolumn{1}{l|}{43.66}                                               &                     \\
\multicolumn{1}{l|}{D=32}                  & 12.04                                                                  & 97.64                                                                   & \multicolumn{1}{l|}{46.77}                                               & 11.27                                                                  & 98.00                                                                   & \multicolumn{1}{l|}{46.96}                                               & 11.48                                                                  & 97.99                                                                   & \multicolumn{1}{l|}{46.81}                                               &                     \\
\multicolumn{1}{l|}{D=64}                  & 10.24                                                                  & 98.29                                                                   & \multicolumn{1}{l|}{47.52}                                               & 9.56                                                                   & 98.54                                                                   & \multicolumn{1}{l|}{47.65}                                               & 10.36                                                                  & 98.17                                                                   & \multicolumn{1}{l|}{47.43}                                               &                     \\
\multicolumn{1}{l|}{D=128}                 & 9.44                                                                   & 98.61                                                                   & \multicolumn{1}{l|}{47.97}                                               & 9.36                                                                   & 98.63                                                                   & \multicolumn{1}{l|}{47.96}                                               & 10.16                                                                  & 98.17                                                                   & \multicolumn{1}{l|}{47.69}                                               &                     \\
\multicolumn{1}{l|}{D=256}                 & 9.90                                                                   & 98.35                                                                   & \multicolumn{1}{l|}{48.18}                                               & 10.49                                                                  & 97.98                                                                   & \multicolumn{1}{l|}{47.99}                                               & 12.59                                                                  & 97.35                                                                   & \multicolumn{1}{l|}{46.91}                                               &                     \\
\multicolumn{1}{l|}{D=512}                 & 10.98                                                                  & 97.80                                                                   & \multicolumn{1}{l|}{47.56}                                               & 10.82                                                                  & 97.81                                                                   & \multicolumn{1}{l|}{47.80}                                               & 11.48                                                                  & 97.88                                                                   & \multicolumn{1}{l|}{47.30}                                               &                     \\
\multicolumn{1}{l|}{D=2048}                & 11.93                                                                  & 97.32                                                                   & \multicolumn{1}{l|}{47.40}                                               & 11.76                                                                  & 97.73                                                                   & \multicolumn{1}{l|}{47.59}                                               & 11.76                                                                  & 97.20                                                                   & \multicolumn{1}{l|}{47.14}                                               &                     \\ \hline
\hline
\multicolumn{11}{c}{\textbf{Our Proposed PFJM}}                                                                                                                                                                                                                                                                                                                                                                                                                                                                                                                                                                                                                                                                                                             \\ \hline
\multicolumn{1}{l|}{D=2}                   & 11.65                                                                  & 97.82                                                                   & \multicolumn{1}{l|}{47.33}                                               & 11.33                                                                  & 97.64                                                                   & \multicolumn{1}{l|}{47.35}                                               & 10.79                                                                  & 97.82                                                                   & \multicolumn{1}{l|}{47.39}                                               & \multirow{8}{*}{10} \\
\multicolumn{1}{l|}{D=8}                   & 10.22                                                                  & 98.11                                                                   & \multicolumn{1}{l|}{47.55}                                               & 10.15                                                                  & 98.17                                                                   & \multicolumn{1}{l|}{47.64}                                               & 10.32                                                                  & 98.08                                                                   & \multicolumn{1}{l|}{47.78}                                               &                     \\
\multicolumn{1}{l|}{D=32}                  & 9.62                                                                   & 98.39                                                                   & \multicolumn{1}{l|}{47.98}                                               & 9.23                                                                   & 98.38                                                                   & \multicolumn{1}{l|}{48.01}                                               & 9.92                                                                   & 98.33                                                                   & \multicolumn{1}{l|}{47.95}                                               &                     \\
\multicolumn{1}{l|}{D=64}                  & 9.11                                                                   & 98.72                                                                   & \multicolumn{1}{l|}{48.20}                                               & 8.99                                                                   & 98.70                                                                   & \multicolumn{1}{l|}{48.17}                                               & 9.52                                                                   & 98.52                                                                   & \multicolumn{1}{l|}{48.03}                                               &                     \\
\multicolumn{1}{l|}{\textbf{D=128}}        & \textbf{8.85}                                                          & \textbf{98.81}                                                          & \multicolumn{1}{l|}{\textbf{48.29}}                                      & \textbf{8.86}                                                          & \textbf{98.81}                                                          & \multicolumn{1}{l|}{\textbf{48.28}}                                      & \textbf{9.28}                                                          & \textbf{98.62}                                                          & \multicolumn{1}{l|}{\textbf{48.15}}                                      &                     \\
\multicolumn{1}{l|}{D=256}                 & 9.26                                                                   & 98.75                                                                   & \multicolumn{1}{l|}{48.22}                                               & 10.17                                                                  & 98.10                                                                   & \multicolumn{1}{l|}{47.93}                                               & 10.68                                                                  & 97.98                                                                   & \multicolumn{1}{l|}{47.58}                                               &                     \\
\multicolumn{1}{l|}{D=512}                 & 10.73                                                                  & 98.08                                                                   & \multicolumn{1}{l|}{47.67}                                               & 10.43                                                                  & 98.21                                                                   & \multicolumn{1}{l|}{47.84}                                               & 10.34                                                                  & 98.07                                                                   & \multicolumn{1}{l|}{48.11}                                               &                     \\
\multicolumn{1}{l|}{D=2048}                & 11.05                                                                  & 98.09                                                                   & \multicolumn{1}{l|}{47.87}                                               & 10.65                                                                  & 98.26                                                                   & \multicolumn{1}{l|}{47.96}                                               & 10.48                                                                  & 98.03                                                                   & \multicolumn{1}{l|}{48.04}                                               &                     \\ 
\toprule[1.5pt]
\end{tabular}
\label{Tab1}
\end{table*}

\section{Experimental Design and Results}
\subsection{Materials and Metrics}
\begin{figure*}[!htb]
    \centering
    \resizebox{1.0\textwidth}{!}
    {\includegraphics{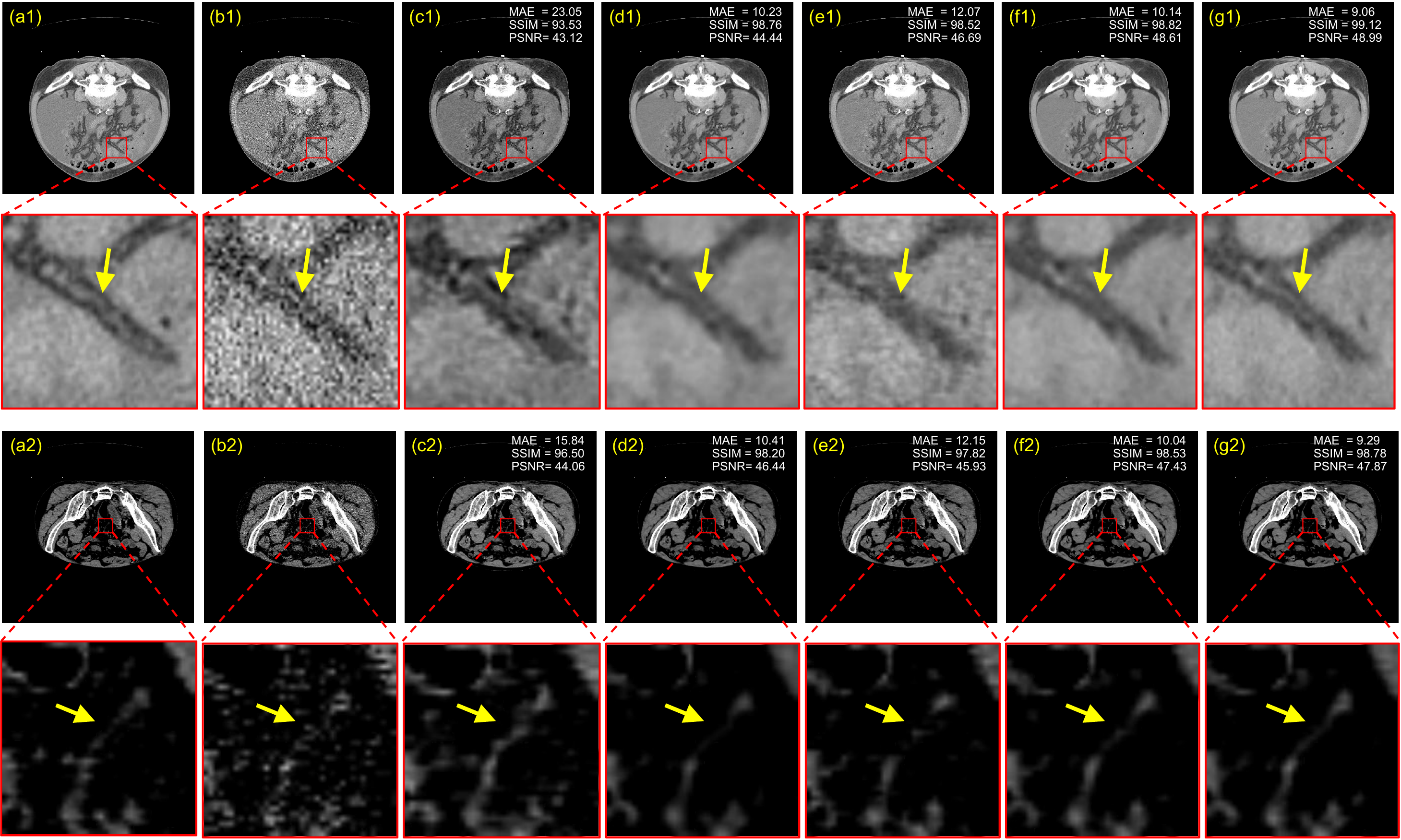}}
    \caption{Qualitative denoising results at Phase I. (a1) and (a2) Routine-dose MCCT images (the ground truth), (b1) and (b2) low-dose CT images, (c1) and (c2) PFGM, (d1) and (d2) DDIM, (e1) and(c2) EDM, (f1) and (f2) PFGM++, (f1) and (f2) our proposed PFJM. The sampling steps for PFGM, DDIM, and EDM are 30, 20 and 20, respectively. Both PFGM++ and our PFJM  use the same number of sampling steps = 10 and $D$ = 128. The display window for (a1)-(g1) is [-150, 150] HU, while the window for the rest images is [-65,205] HU. The red-boxed ROIs are zoomed for detailed visual comparison.
           }
    \label{result1}
\end{figure*}

\begin{figure*}[!htb]
    \centering
    \resizebox{1.0\textwidth}{!}
    {\includegraphics{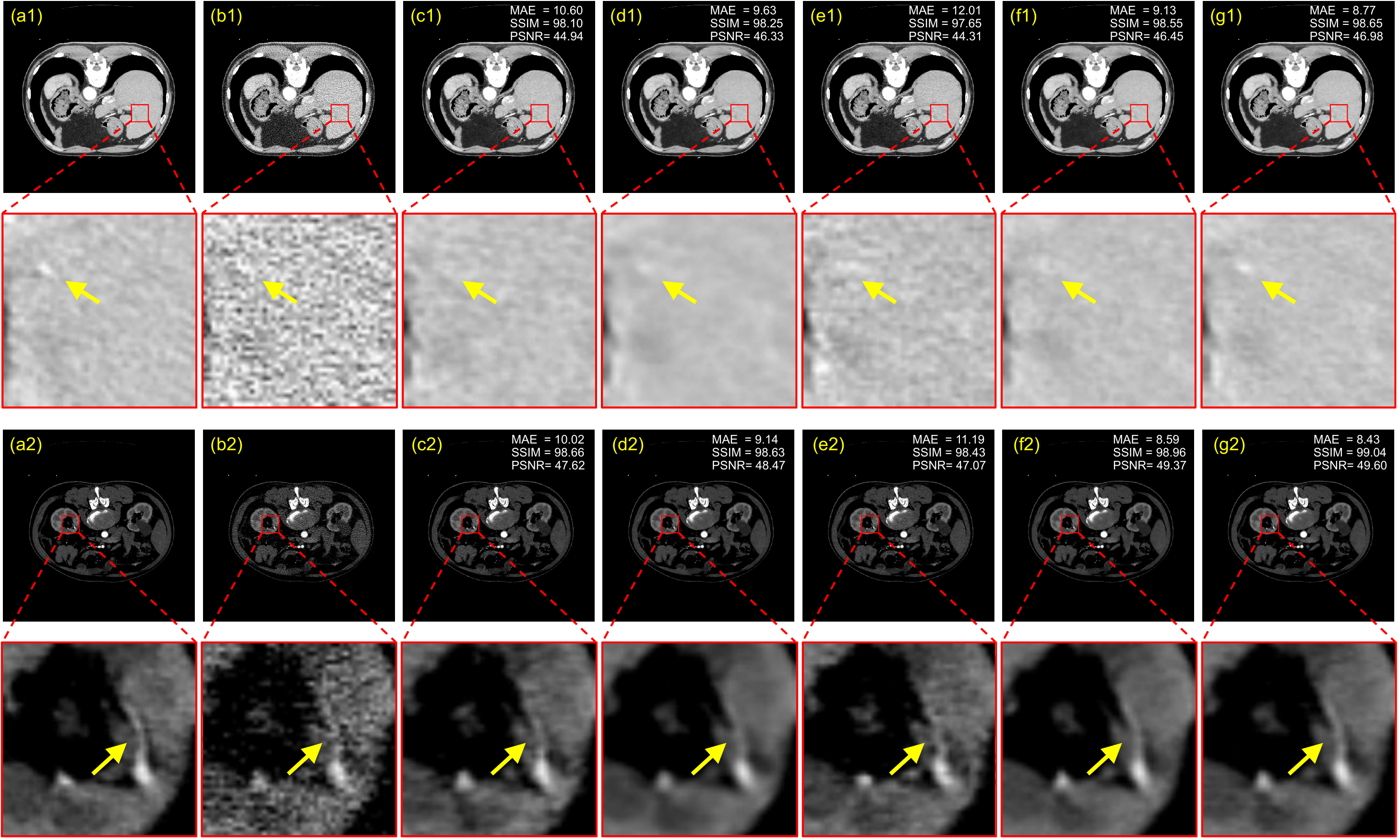}}
    \caption{ Qualitative denoising results at Phase II. (a1) and (a2) Routine-dose CECT images (the ground truth), (b1) and (b2) low-dose MCCT images, (c1) and (c2) PFGM, (d1) and (d2) DDIM, (e1) and (c2) EDM, (f1) and (f2) PFGM++, (f1) and (f2) our proposed PFJM. The sampling steps for PFGM, DDIM, and EDM are 30, 20 and 20, respectively. Both PFGM++ and our PFJM use the same number of sampling steps = 10 and $D$ = 128. The display window for (a1)-(g1) is [-150, 150] HU, while the window for the rest images is [-50,350] HU. The red-boxed ROIs are zoomed for detailed visual comparison.
           }
    \label{result2}
\end{figure*}

\textbf{Dataset.} In this study,112,092 MCCT images from 131 patients are collected from the public dataset ``VinDr-Multiphase'' \cite{VinDr}. Each patient was CT-scanned during non-contrast, arterial, and venous phases respectively. 
The dataset was randomly divided into 65 patients’MCCT scans for training, 21 for validation, and 45 for testing, with no data overlap. The corresponding low-dose MCCT images were simulated, with the effects of the bowtie filter, automatic exposure control, and electronic noise \cite{Scan_protocol}. Specifically, we simulated $10\%$ dose data that bear remarkable image noise in all phases.   
Visually as (f1) and (f2) in Figs.~\ref{result1}-\ref{result3}, it comprehensively enables both superior noise suppression and detail preservation in all phases, with performance close to the routine-dose MCCT images. 

\textbf{Evaluation Metrics.} The deep denoising performance of the models was assessed in terms of the following popular metrics: 
\begin{itemize}
\item 
The Mean Absolute Error (MAE) defined as 
\begin{equation}
\text{MAE} = \frac{1}{N} \sum_{i=1}^{N} |I_{\text{RD}}(i) - I_{\text{RECON}}(i)|,
\label{MAE}
\end{equation}
where $I_{\text{RD}}(i)$ and $I_{\text{RECON}}(i)$ are the voxel values at position $i$ in routine-dose and reconstructed images.
\item 
The Structural Similarity (SSIM) defined as
\begin{equation}
\text{SSIM} = \frac{(2\mu_{_{\text{RD}}} \mu_{_{\text{RECON}}} + C_1)(2\sigma_{_{\text{RD} \cdot \text{RECON}}} + C_2)}{(\mu_{_{\text{RD}}}^2 + \mu_{_{\text{RECON}}}^2 + C_1)(\sigma_{_{\text{RD}}}^2 + \sigma_{_{\text{RECON}}}^2 + C_2)},
\label{SSIM}
\end{equation}
where $\mu_{_{\text{RD}}}$ and $\mu_{_{\text{RECON}}}$ are the average voxel values of the routine-dose and reconstructed images, $\sigma_{_{\text{RD}}}^2$ and $\sigma_{_{\text{RECON}}}^2$ represent the corresponding variances, and $\sigma_{_{\text{RD} \cdot \text{RECON}}}$ is the covariance between $I_{\text{RD}}$ and $I_{\text{RECON}}$.
\item 
The Peak Signal-to-Noise-Ratio (PSNR) defined as
\begin{equation}
\text{PSNR} = 10 \cdot \log_{10} \left( \frac{MAX^2}{\text{MSE}} \right),
\label{PSNR}
\end{equation}
where $MAX$ is the maximum pixel value, and $MSE$ denotes the Mean Squared Error between $I_{\text{ND}}$ and $I_{\text{RECON}}$.
\item 
The Fréchet Inception Distance (FID) score defined as
\begin{equation}
\text{FID} = \left \| \mu_{_{r}} - \mu_{_{g}} \right \| _2^2+Tr({\sum}_{r}+{\sum}_{g}-2({\sum}_{r}{\sum}_{g})^{1/2} ),
\label{FID}
\end{equation}
where $\mu_{_{r}}$ and ${\sum}_{r}$ are the mean and covariance matrices of the routine-dose image features, respectively, while $\mu_{_{g}}$ and ${\sum}_{g}$ are used for the reconstructed image features.
\end{itemize}

\begin{table}[t]
\centering
\caption{FID score for reconstruction quality comparison}
\begin{tabular}{ccccc}
\bottomrule[1.5pt]
PFGM          & DDIM          & EDM           & PFGM++               & \textbf{PFJM }                \\
{[}30 step{]} & {[}20 step{]} & {[}20 step{]} & $\left[\begin{array}{c}\text{10 step}\\ \text{D=128}\end{array}\right]$  & $\left[\begin{array}{c}\text{10 step}\\ \text{D=128}\end{array}\right]$ \\ \hline
27.23         & 10.87         & 35.90         & 6.34                 & \textbf{4.30 }                \\ 
\toprule[1.5pt]
\end{tabular}
\label{FID}
\end{table}

\subsection{Results and Analysis}
\begin{figure*}[!htb]
    \centering
    \resizebox{1.0\textwidth}{!}
    {\includegraphics{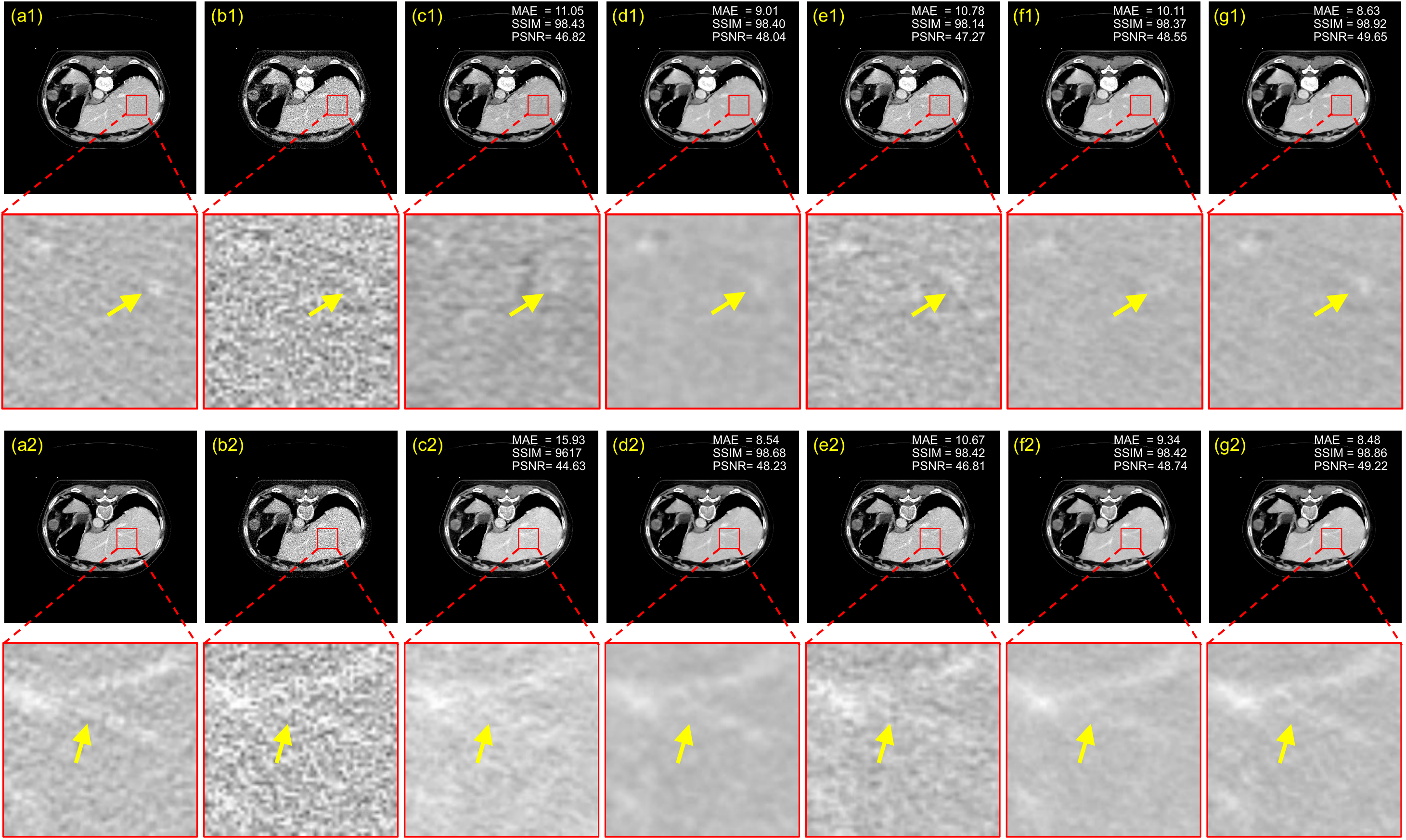}}
    \caption{Qualitative denoising results at Phase III. (a1) and (a2) Routine-dose MCCT images (the ground truth), (b1) and (b2) low-dose MCCT images, (c1) and (c2) PFGM, (d1) and (d2) DDIM, (e1) and (c2) EDM, (f1) and (f2) PFGM++, (f1) and (f2) our proposed PFJM. The sampling steps for PFGM, DDIM, and EDM are 30, 20 and 20, respectively. Both PFGM++ and our PFJM use the same number of sampling steps = 10 and $D$ = 128. The display window is [-100, 200] HU. The red-boxed ROIs are zoomed for detailed visual comparison.
           }
    \label{result3}
\end{figure*}

\textbf{Overall performance.} 
As shown in Table.~\ref{Tab1}, our proposed PFJM achieved superior 
MCCT denoising results, with MAE down to 8.99 HU, SSIM up to 98.75\%, and PSNR up to 46.3 dB, averaged over phases I, II and III.
Visually as (f1) and (f2) in Figs.~\ref{result1}-\ref{result3}, it comprehensively enables both superior noise suppression and detail preservation in all phases, with performance close to the routine-dose MCCT images. 
The results demonstrate the proposed PFJM outperformed the competing models: 1) PFGM \cite{PFGM}, 2) diffusion models DDIM\cite{DDIM} and EDM\cite{Karras2022}, as well as 3) PFGM++. All of these competing models are updated with the multi-phase joint condition in network learning same as PFJM, avoiding unconditional generation that is not applicable.  
All of these competing models are updated with the multi-phase joint condition in network learning same as PFJM, avoiding unconditional generation that is not applicable.

With just 10 sampling steps, the proposed PFJM ($D = 128$) already surpassed PFGM, DDIM, and EDM that took 30, 20, and 20 steps respectively. PFGM even failed to reconstruct decent images when the number of sampling steps are less than 30, because of its demand of extra large batches to construct the electric field.  
Given the same number of steps used by PFJM, which is 10 steps, both DDIM and EDM showed performance degradation, while PFJM outperformed them clearly, decreasing MAE by $5.06$ HU, increasing SSIM and PSNR by $1.70\%$ and $3.51$db on average. 

Although PFGM++ had worse performance than the other compared models with augmented dimension $D = \{2,8,32\}$, it can adapt the task gradually by changing $D$ to improve results. 
Our PFJM still suppresses PFGM++ at each $D$, as shown in Table.~\ref{Tab1}.
It is also feasibly adjustable by D to adaptively
optimize multiphase joint reconstruction path, with performance enhancements.
Furthermore, the multi-phase joint condition helped in not only network learning but also generative sampling such that our PFJM can outperform PFGM++ even at $D=2$, with MAE decreased by $49.17$ HU, SSIM increased by $27.58\%$, and PSNR increased by $35.35$dB.

Furthermore, FID score was calculated across all phases to reveal superiority of our proposed PFJM in reconstruction quality, as shown in Table.~\ref{FID}. Reconstruction of three phases was taken as three channels of image, so that FID can evaluate the joint distribution of multiphase results compared with the real MCCT data. Obviously, our PFJM achieved the best FID score of $4.30$, meaning its multiphase reconstruction had extremely high realism. It effectively decreased the FID score by $15.79$ in average, compared to PFGM, DDIM, EDM, and PFGM++. It was demonstrated that our PFJM was good at learning the jointly mapping from the low-dose MCCT images to the routine-dose.

\textbf{Analysis on Phase I reconstruction.}
Fig.~\ref{result1} shows qualitative denoising results at Phase I, and the region of interest (ROI) is zoomed in for detailed visualization. MCCT at phase I acquired without any contrast serves as the structural baseline. 
The low-dose result in Fig.~\ref{result1}(b1) exhibits that it suffers from serious noise, and its mesenteric vessel structure in the enlarged ROI is significantly damaged, unacceptable for clinical diagnosis. The PFGM result in Fig.~\ref{result1}(c1) shows its vessel well reconstructed with adjacent tissues as the yellow arrow indicated. 
EDM in Fig.~\ref{result1}(e1) introduces noticeable artifacts and breaks the structure. The DDIM and PDFM++ results in Figs.~\ref{result1}(d1) and (f1) seem excessively denoised so that the partial structures are diluted. The proposed PFJM in Fig.~\ref{result1}(g1) is the best result with noise effectively suppressed and structure clearly maintained. 
Furthermore, for the mesenteric vessel in the pelvic region in Fig.~\ref{result1}(a2), the low-dose result (Fig.~\ref{result1}(b2)) fails to capture key features under excessive noise. PFGM is confused by noise, compromises vessel shape, and alters pixel HU value, as shown in Fig.~\ref{result1}(c2). Although diffusion models DDIM and EDM in Figs.~\ref{result1}(d2) and (e2) effectively suppress the noise, the vessel is broken or lost. PFGM++ and the proposed PFJM in Figs.~\ref{result1}(f2) and (g2) both show superiority in detailed reconstruction and denoising effects. Thanks to the generation direction refinement of sampling in PFJM, the vessel structure is faithfully recovered, as indicated by arrows.

Quantitatively, in terms of the metrics used for Phase I in Table.~\ref{Tab1}, our PFJM with $D=128$ produced the best performance with low MAE of $8.85$HU, high SSIM of $98.81$\%, and high PSNR of $48.29$db. It outperformed the competing methods PFGM (step=30), DDIM (step=\{10,20\}), EDM (step=\{10,20\}), and PFGM++ (step=10, $D$=128), with MAE decreased by $3.25$ HU, SSIM increased by $1.12$\%, and PSNR increased by $2.36$db.

\begin{figure*}[t]
    \centering
    \subfigure[Influence of $D$ on MAE ]{\includegraphics[width=0.31\textwidth]{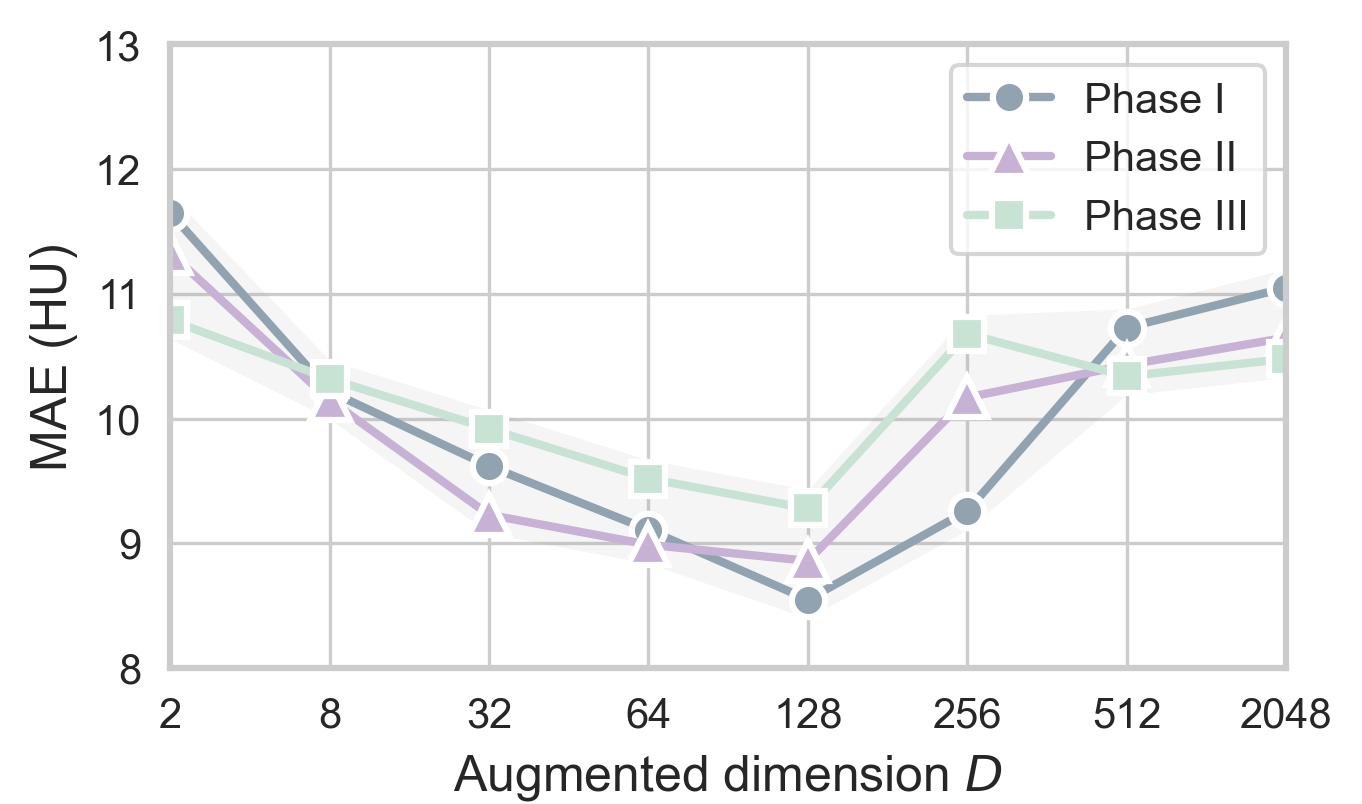}} 
    \hspace{0.1cm} 
    \subfigure[Influence of $D$ on SSIM]{\includegraphics[width=0.31\textwidth]{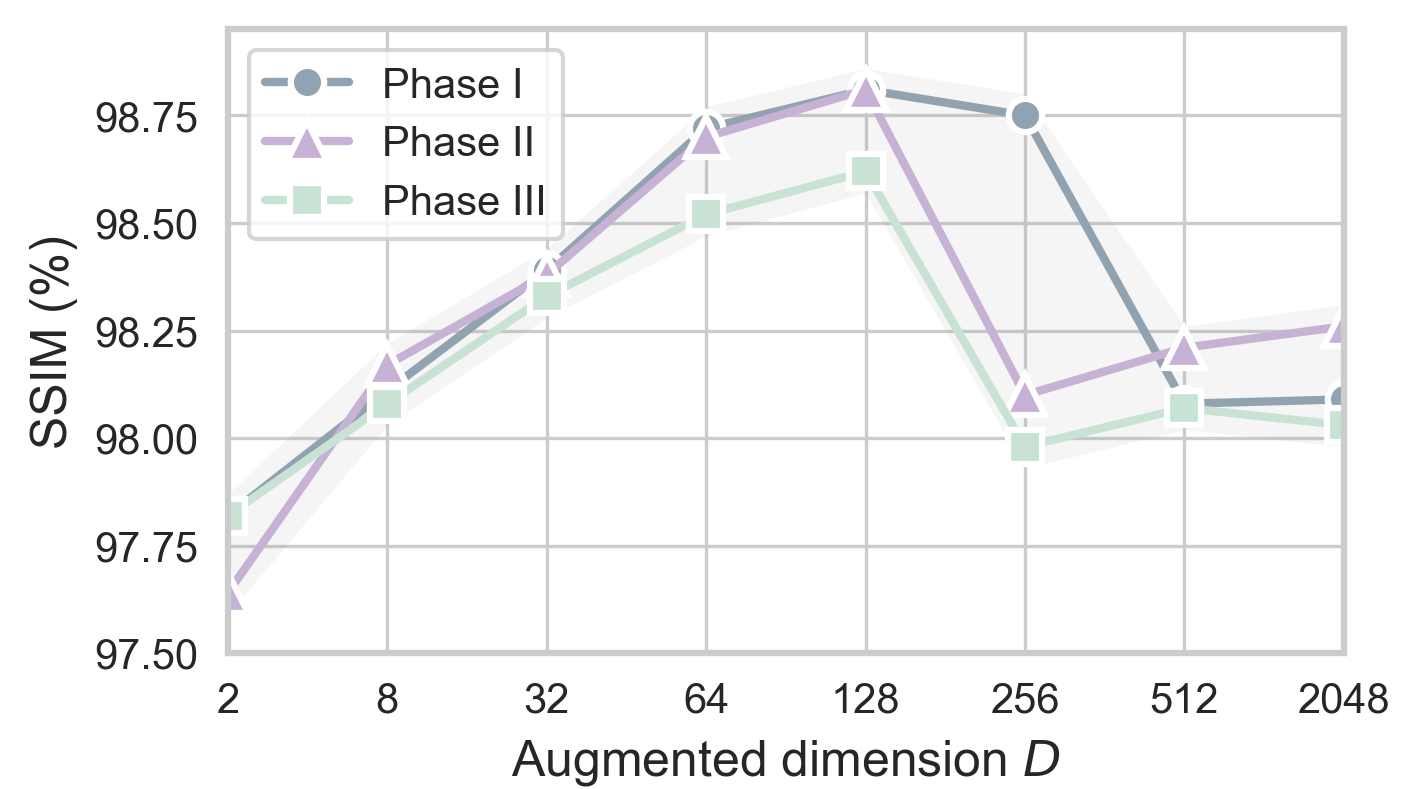}}
    \hspace{0.1cm} 
    \subfigure[Influence of $D$ on PSNR]{\includegraphics[width=0.31\textwidth]{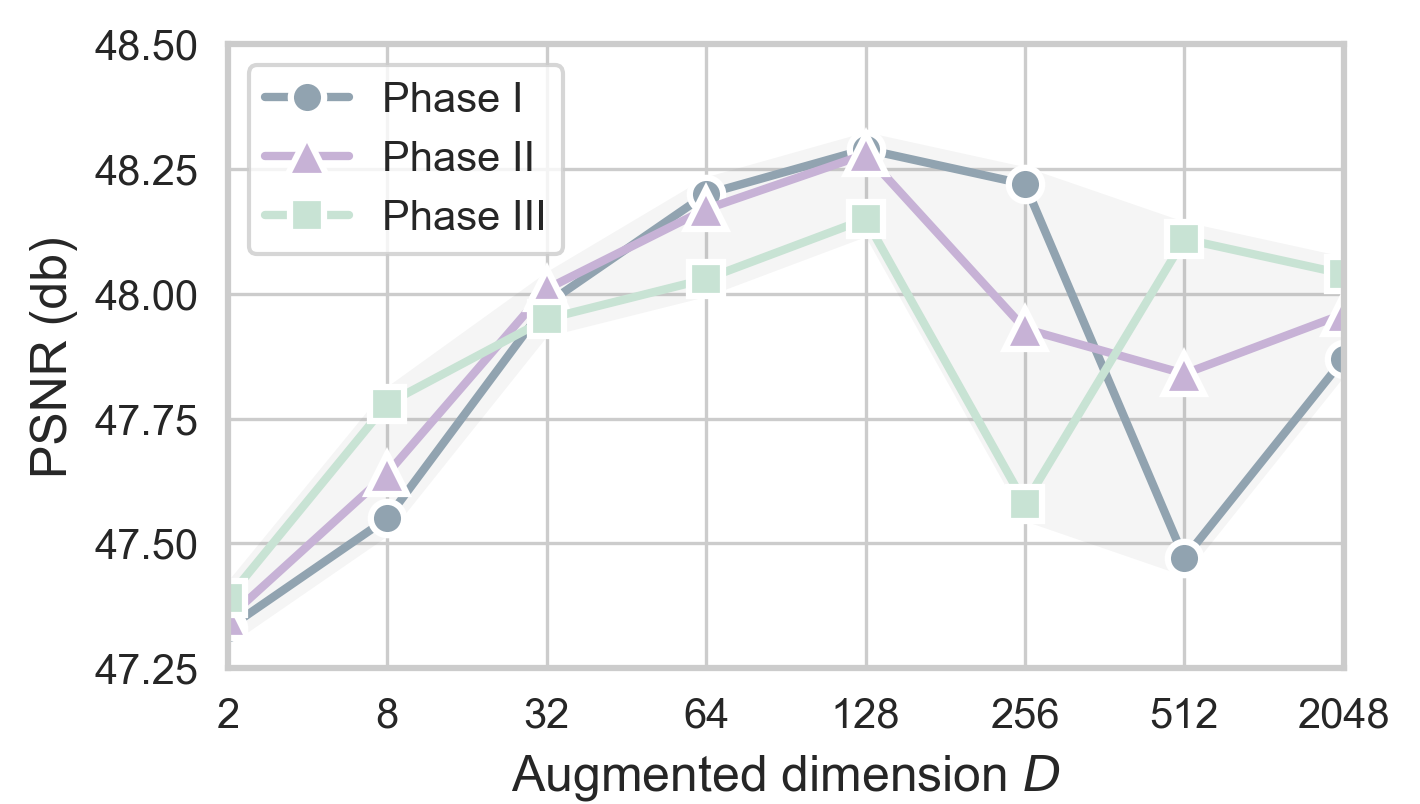}} 
    \caption{Charts demonstrating the influence of various $D$ selections in PFJM for MCCT imaging. Notably, $D=128$ achieves the best performance consistently across all metrics and phases as the optimal selection for our specific task.}
    \label{trend}
\end{figure*}

\textbf{Analysis on Phase II reconstruction.}
Fig.~\ref{result2} presents an impression on the denoising results at Phase II, with the ROI enlarged for details. MCCT at phase II is acquired after the contrast agent flows into artery, highlighting the vascular for diagnosis. 
For the low-dose result shown in Fig.~\ref{result2}(b1), the small hepatic artery branch cannot be distinguished due to the severe noise. PFGM in Fig.~\ref{result2}(c1) loses this vessel due to noise. For DDIM in Fig.~\ref{result2}(d1), although the vessel is preserved, its clarity is low due to over-denoising. The small vessel structure in cross-section reconstructed by EDM is evidently inaccurate because of the additional artifacts in Fig.~\ref{result2}(e1). PFGM++ in Fig.~\ref{result2}(f1) has shown its effectiveness in keeping 
structural fidelity, but its contrast is low, such as the small vessel indicated by the arrow is easily overlooked. The proposed PFJM achieves the best details and denoising effect as shown in Fig.~\ref{result2}(g1), which is remarkably consistent with the ground truth. The small arterial vessel is precisely reconstructed with  clear contrast to the surrounding.    
Furthermore, for the renal artery branch in Fig.~\ref{result2}(a2), the strong noise in the low-dose results (Fig.~\ref{result2}(b2)) covers the structural details in the MCCT image.  The PFGM result in Fig.~\ref{result2}(c2) illustrates that the partial artery structure is erroneously eliminated along with noise, and the rest looks confusing with the kidney. DDIM reduced the noise in MCCT, but structural details such as the renal artery in ROI show contrast agent dissipation and attenuation inconsistent with the ground truth in Fig.~\ref{result2}(d2). EDM distorted vessel shape, as indicated by the arrow in Fig.~\ref{result2}(e2). PFGM++ made progress compared to the PFGM and diffusion models, in denoising and restoration. Most notably, the proposed PFJM in Fig.~\ref{result2}(g2) achieved much better details for vessel contours and so on with contrast 
enhancement, being the same as the ground truth. Thus, it improves the anatomical visualization of both small details and overall artery configuration with low-dose MCCT.

In the quantitative evaluation, the proposed PFJM with $D=128$ surpassed the competing Poission flow models and diffusion models for MCCT at Phase II, as shown in Table.~\ref{Tab1}. On average, it produced a $2.98$HU $\downarrow$ in MAE, a $1.01\%$ $\uparrow$ in SSIM, and a $2.23$db $\uparrow$.

\begin{figure}[t]
    \centering
    \resizebox{0.7\textwidth}{!}
    {\includegraphics{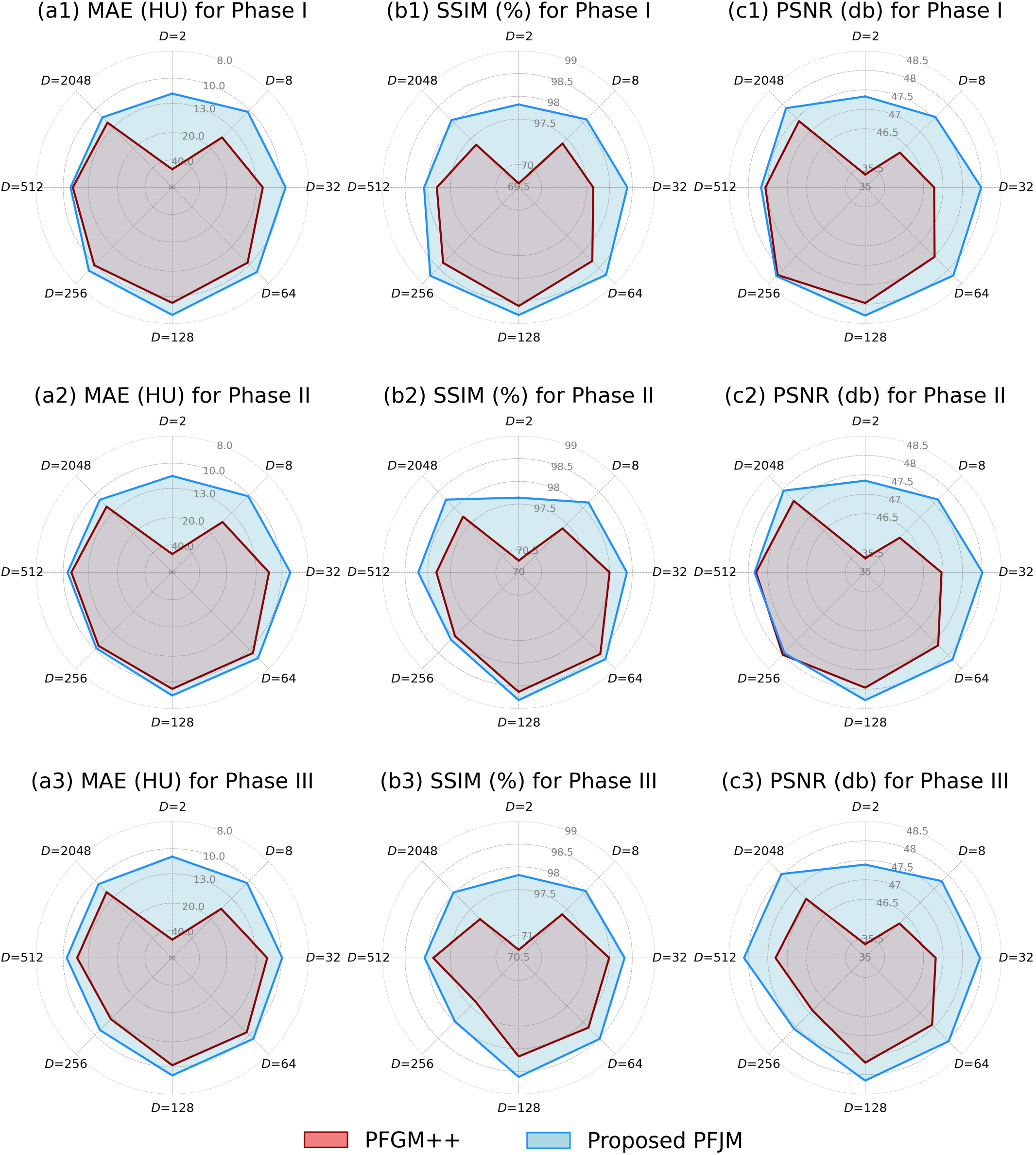}}
    \caption{Radar charts comparing the performance with and without the conditional sampling, i.e., PFJM and PFGM++,  across different $D$ in terms of MAE (a1)-(a3), SSIM (b1)-(b3), and PSNR (c1)-(c3). The charts demonstrate the advantages of the conditional sampling in enhancing reconstruction quality by optimizing the dimensionality variable $D$.
           }
    \label{rada}
\end{figure}

\textbf{Analysis on Phase III reconstruction.}
The qualitative denoising results of Phase III are exhibited in Fig.~\ref{result3}. MCCT at phase III was acquired when the contrast agent enters into the venous system to visualize veins. 
For the hepatic vein branch in Fig.~\ref{result3}(a1), it is difficult to observe in the low-dose result (Fig.~\ref{result3}(b1)) given the heavy noise. The PFGM and EDM results suffer from low signal-to-noise ratios, disallowing clear distinction of the venous vessel, as shown in Figs.~\ref{result3}(c1) and (e1). Due to the severe noise, DDIM and PFGM++ denoised this small structure but weakened its contrast and diluted the signal, as shown in Figs.~\ref{result3}(d1) and (f1). The proposed PFJM successfully preserved the fine venous structure and suppressed the noise, as shown by the arrow in Figs.~\ref{result3}(e1). 
Furthermore, for the hepatic vein region in Fig.~\ref{result3}(a2), the low-dose result in Fig.~\ref{result3}(b2) contained a significantly high noise level, compromising the anatomical structure. PFGM reduced the noise, but the edge details were compromised, and the vein's contour remained still blurry, as shown in Fig.~\ref{result3}(b2). DDIM and PFGM++ in Figs.~\ref{result3}(d2) and (f2) made better noise suppression than PFGM, but vein edge blurring persisted with a decreased contrast. EDM in Fig.~\ref{result3}(e2) shows the loss of fine details, such as the vein boundary, and demonstrates limited structural preservability. As shown in Fig.~\ref{result3}(f2), the proposed PFJM achieved the best noise suppression among all methods and  the most faithful anatomical features, providing diagnostic quality. It is an excellent balance between denoising and preserving fine details, closely matching the routine-dose image.

As shown in Table~\ref{Tab1}, the quantitative results demonstrate that the proposed PFJM with $D=128$ achieves the best performance across all metrics, with the lowest MAE (9.28HU), the highest SSIM (98.62\%), and the highest PSNR (48.15db), for MCCT at Phase III. On average, the gains include $2.92$HU $\downarrow$ in MAE, $0.97\%$ $\uparrow$ in SSIM, and $2.32$db $\uparrow$ in PSNR, compared to the competing methods.

\begin{figure}[t]
    \centering
    \resizebox{0.9\textwidth}{!}
    {\includegraphics{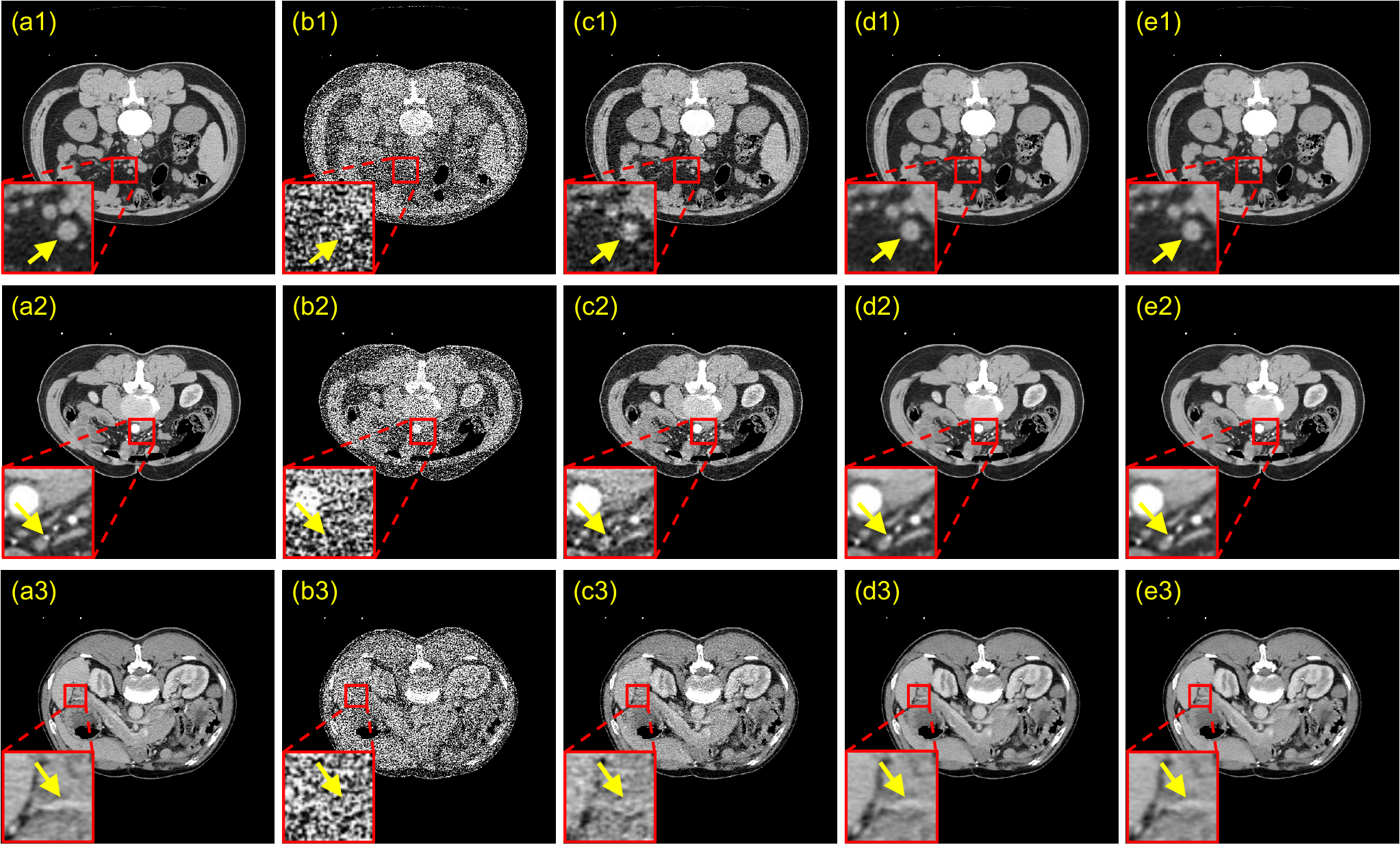}}
    \caption{Qualitative comparison of the proposed PFJM between with and without the conditional sampling, across all phases under small $D=\{2,8\}$. (a1)-(a3) Routine-dose MCCT images (the ground truth) at Phase I, II and III, respectively. Correspondingly, (b1)-(b3) the results \textit{w/o} conditional sampling [D=2], (c1)-(c3) results \textit{w/o} conditional sampling [D=8], (d1)-(d3) results \textit{w/} conditional sampling [D=2], and (e1)-(e3) results \textit{w/} conditional sampling [D=8]. The display window for (a1)-(e1) and (a2)-(e3) is [-150, 150] HU, while the window for (a3)-(e3) is [-100, 200] HU. The red-boxed ROIs are zoomed for visual comparison. By conditional sampling, the structural details are preserved, making the results consistent with routine-dose MCCT images, even for the ultra small $D$ that leads to unacceptable reconstruction without this strategy.
           }
    \label{ablation}
\end{figure}

\textbf{Ablation study on the augmented dimension $\bm{D}$.}
To make the optimal selection of the augmented dimension $D$, the settings experimented include $D=\{2,8,32,,64,128,,256,512,2048\}$. As illustred in Table~\ref{Tab1}, the performance can be adjusted with $D$, to adapt  the MPCT joint reconstruction for the task of our interest. $D=128$ achieved the best performance among these settings, for phases I, II and III. To analyze the trends, Fig.~\ref{trend} displays the continuous changes in MAE, SSIM, and PSNR  with respect to different $D$ values. For the range before $D=128$, the performance improves rapidly as $D$ increases. For the range after $D=128$, the performance decreases, but despite large changes in D values, the rate of performance variation is slower than that in the former range. As the sweet spot, $D=128$ represents a excellent trade-off between robustness and rigid, as it results in optimal coupling between the data and the additional variable norms.

\textbf{Ablation study on the effect of conditional sampling.} PFJM was further evaluated in the situation where the the multi-phase joint condition was removed in generative sampling, i.e. PFJM was degraded to PFGM++. The comparison between PFJM and PFGM++ in Table~\ref{Tab1}, shows that the conditional sampling improved the imaging performance with $7.58$HU decrement in MAE, $3.97\%$ increment in SSIM, and $2.40$db increment in PSNR, averaged on all $D$ selections and phases. As shown in Fig.~\ref{rada}, the conditional sampling explicitly empowered PFJM to outperform PFGM++ on each $D$ selection for all phases I, II, and III. Especially, it remarkably improved the reconstruction even for small $D$ values of $\{2,8\}$ when PFGM++  failed or performed poor. As Fig.~\ref{ablation} demonstrates, although the small values $D=\{2,8\}$ led to fully unacceptable reconstructions using PFGM++,  the conditional sampling still successfully fixed these issues even with the same small $D$ values. This conditional strategy enables the structural details consistent with that in the routine-dose MCCT images across all phases in Fig.~\ref{ablation}(a1)-(a3). Then, the effect of conditional sampling with the best $D=128$ can be found in Figs.~\ref{result1}-\ref{result3} (f1)\&(f2) vs. (g1)\&(g2). It further refined the  anatomical visualization for MCCT, improving structural details, contrast, clarity, discriminability and integrity, because the generation direction was adjusted at each sampling step with the prior joint condition.


\section{Discussions and Conclusion}
Leveraging the PFGM++ architecture, the proposed model reformulates the reconstruction task into learning the joint distribution mapping to the routine-dose MCCT images. By treating data distributions as electric charges in an augmented space with dimensionality $N+D$, PFJM allows the adjustment of the augmented dimension $D$, facilitating adaptive optimization of the generation path. This innovation enables the reconstruction process to achieve high-quality imaging with minimized radiation exposure in MCCT imaging.

Another key contribution of this work is the design of the multi-phase joint condition into both network learning and generation sampling. This mechanism effectively utilizes prior distributions of correlated anatomical structures across phases, accelerating the convergence of the generative trajectory toward the posterior distribution of routine-dose MCCT images. Its conditional sampling process further refines the generation direction iteratively, ensuring superior recovery of anatomical fidelity. Extensive experiments produced promising results, including average MAE down to $8.99$HU, SSIM up to $98.75\%$, and PSNR up to $48.24$db. These results suggest a significant clinical potential of PFJM in reducing the radiation dose and improving diagnostic performance of MCCT examinations. 

In conclusion, we have proposed a novel PFJM model for low-dose MCCT, which is the first framework based on ``PFGM++'' for low-dose MCCT imaging. It suppresses image noise caused by photon starvation in low-dose MCCT, and at the same preserves clinical features with sufficient details. In our follow-up work, we plan to use diverse big data, address generalizability and fairness, and translate into clinical practice.


\begin{thebibliography}{1}

\bibitem{Meng2020}
X.P.Meng, Y.C.Wang, S.Ju, et al., \emph{Radiomics analysis on multiphase contrast-enhanced CT: a survival prediction tool in patients with hepatocellular carcinoma undergoing transarterial chemoembolization}, Frontiers in Oncology, 2020, 10: 1196.

\bibitem{Smithuis}
R. ~Smithuis, \emph{CT contrast injection and protocols. Radiology Assistant}, https://radiologyassistant. nl/more/ct-protocols/ct-contrast-injection-and-protocols, 2014.

\bibitem{Brenner}
D.J.~Brenner, E.J.~Hall, \emph{Computed tomography an increasing source of radiation exposure}, New England journal of medicine, 2007, 357(22): 2277-2284.

\bibitem{Rastogi}
S.~Rastogi, R.~Singh, R.~Borse, et al., \emph{Use of multiphase CT protocols in 18 countries: appropriateness and radiation doses}, Canadian Association of Radiologists Journal, 2021, 72(3): 381-387.

\bibitem{Prasad}
K.N.~Prasad, W.C.Cole, G.M.~Haase, \emph{Radiation protection in humans: extending the concept of as low as reasonably achievable (ALARA) from dose to biological damage}, The British journal of radiology, 2004, 77(914): 97-99.


\bibitem{RED_CNN}
H.~Chen, Y.~Zhang, M. K.~Kalra, F.~Lin, Y.~Chen, P.~Liao,
J.~Zhou, G.~Wang, \emph{Low-dose CT with a residual encoder-decoder convolutional neural network}, IEEE transactions on medical imaging, 2017, 36(12): 2524-2535.

\bibitem{Yin2019}
X.~Yin, Q.~Zhao, J.~Liu, et al., \emph{Domain progressive 3D residual convolution network to improve low-dose CT imaging}, IEEE transactions on medical imaging, 2019, 38(12): 2903-2913.

\bibitem{Fan2019}
F.~Fan, H.~Shan, M.K.~Kalra, et al., \emph{Quadratic autoencoder (Q-AE) for low-dose CT denoising}, IEEE transactions on medical imaging, 2019, 39(6): 2035-2050.

\bibitem{Lu2022}
Z.~Lu, W.~Xia, Y.~Huang, et al., \emph{$M_3$NAS: Multi-scale and multi-level memory-efficient neural architecture search for low-dose ct denoising}, IEEE Transactions on Medical Imaging, 2022, 42(3): 850-863.

\bibitem{Yang2022}
L.~Yang, Z.~Li, R.~Ge, et al., \emph{Low-dose CT denoising via sinogram inner-structure transformer}, IEEE transactions on medical imaging, 2022, 42(4): 910-921.

\bibitem{Yang2018}
Q.~Yang, P.~Yan, Y.~Zhang, H.~Yu, Y.~Shi, X.~Mou, M.K.~Kalra, Y.~Zhang, L.~Sun, G.~Wang, \emph{Low-dose CT image denoising using a generative adversarial network with Wasserstein distance and perceptual loss}, IEEE transactions on medical imaging, 37(6):1348-1357, 2018.

\bibitem{Bera2021}
S.~Bera, P.K.~Biswas, \emph{Biswas Noise conscious training of non local neural network powered by self attentive spectral normalized Markovian patch GAN for low dose CT denoising}, IEEE Transactions on Medical Imaging, 2021, 40(12): 3663-3673.

\bibitem{Kwon2021}
T.~Kwon, J.C.~Ye, \emph{Cycle-free CycleGAN using invertible generator for unsupervised low-dose CT denoising}, IEEE Transactions on Computational Imaging, 2021, 7: 1354-1368.

\bibitem{Moghari2021}
M.D.~Moghari, L.~Zhou, B.~Yu, et al. \emph{Efficient radiation dose reduction in whole-brain CT perfusion imaging using a 3D GAN: performance and clinical feasibility}, Physics in Medicine \& Biology, 2021, 66(7): 075008.

\bibitem{Shan2018}
H.~Shan, Y.~Zhang, Q.~Yang, et al., \emph{3-D convolutional encoder-decoder network for low-dose CT via transfer learning from a 2-D trained network}, IEEE transactions on medical imaging, 2018, 37(6): 1522-1534.

\bibitem{Huang2021}
Z.~Huang, J.~Zhang, Y.~Zhang, H.~Shan, \emph{DU-GAN: Generative adversarial networks with dual-domain U-Net-based discriminators for low-dose CT denoising}, IEEE Transactions on Instrumentation and Measurement, 71:1-12, 2021.

\bibitem{Ge2019}
R.~Ge, G.~Yang, C.~Xu, Y.~Chen, L.~Luo, S.~Li, \emph{Stereo-correlation and noise-distribution aware ResVoxGAN for dense slices reconstruction and noise reduction in thick low-dose CT}, In International Conference of Medical Image Computing and Computer Assisted Intervention–MICCAI, pp. 328-338, 2019.

\bibitem{Fu2023}
Y.~Fu, S.~Dong, M.~Niu, et al., \emph{AIGAN: Attention–encoding Integrated Generative Adversarial Network for the reconstruction of low-dose CT and low-dose PET images}, Medical Image Analysis, 2023, 86: 102787.

\bibitem{GAN}
I.~Goodfellow, J.~Pouget-Abadie, M.~Mirza, B.~Xu, D.~Warde-Farley, S.~Ozair, A.~Courville, Y.~Bengio, \emph{Generative adversarial nets}, Advances in neural information processing systems, pp. 2672-2680, 2014.


\bibitem{Salimans}
T.~Salimans, I.~Goodfellow, W.~Zaremba, V.~Cheung, A.~Radford, X.~Chen, \emph{Improved techniques for training gans}, Advances in Neural Information Processing Systems, pp. 2226-2234, 2016.

\bibitem{Metz} 
L.~Metz, B.~Poole, D.~Pfau, J.~Sohl-Dickstein, \emph{Unrolled Generative Adversarial Networks}, 5th International Conference on Learning Representations, 2017.

\bibitem{Zhao_empirical} 
S.~Zhao, H.~Ren, A.~Yuan, J.~Song, N.~Goodman, S.~Ermon, \emph{Bias and generalization in deep generative models: An empirical study}, Advances in Neural Information Processing Systems, 2018.

\bibitem{Gao_TMI2023} 
Q.~Gao, Z.~Li, J.~Zhang, Y.~Zhang, H.~Shan, \emph{CoreDiff: Contextual error-modulated generalized diffusion model for low-dose CT denoising and generalization}, IEEE Transactions on Medical Imaging, 43(2):745-759, 2024.

\bibitem{Song2020}
Y.~Song, S.~Ermon, \emph{Improved techniques for training score-based generative models}, Advances in neural information processing systems, 33: 12438-12448, 2020.



\bibitem{Karras2022}
T.~Karras, M.~Aittala, T.~Aila, S.~Laine, \emph{Elucidating the design space of diffusion-based generative models}, Advances in neural information processing systems, 35: 26565-26577, 2022. 

\bibitem{DDIM}
J.~Song, C. Meng, S.~Ermon, Denoising Diffusion Implicit Models. In International Conference on Learning Representations.

\bibitem{PFGM}
 Y.~Xu, Z.~Liu, M.~Tegmark, T.~Jaakkola, \emph{Poisson flow generative models}, Advances in Neural Information Processing Systems, 35, pp.16782-16795, 2022.

\bibitem{PFGMPP}
Y.~Xu, Z.~Liu, Y.~Tian, S.~Tong, M.~Tegmark, T.~Jaakkola, \emph{Pfgm++: Unlocking the potential of physics-inspired generative models}, In International Conference on Machine Learning, pp. 38566-38591, 2023.


\bibitem{Ho2020nips}
J.~Ho, A.~Jain, P.~Abbeel, \emph{Denoising diffusion probabilistic models}, Advances in neural information processing systems, 33: 6840-6851, 2020.

\bibitem{Nichol2021PMLR}
A.Q.~Nichol, P.~Dhariwal, \emph{Improved denoising diffusion probabilistic models}, International conference on machine learning, 2021: 8162-8171.

\bibitem{Song2021ICLR}
J.~Song, C.~Meng, S.~Ermon, \emph{Denoising diffusion implicit models}, in International Conference on Learning Representations, 2021, pp. 1–22


\bibitem{Zhao2023}
W.~Zhao, Y.~Rao, W.~Shi, et al., \emph{Diffswap: High-fidelity and controllable face swapping via 3d-aware masked diffusion}, Proceedings of the IEEE/CVF Conference on Computer Vision and Pattern Recognition, pp. 8568-8577, 2023.

\bibitem{Song2021}
Y.~Song, J.~Sohl-Dickstein,D.P.~Kingma, et al., \emph{Score-Based Generative Modeling through Stochastic Differential Equations}, International Conference on Learning Representations, 2021.

\bibitem{Saharia2022}
C.~Saharia, W.~Chan, H.~Chang, et al., \emph{Palette: Image-to-image diffusion models}, in Proceedings of ACM SIGGRAPH Conference, pp. 1-10, 2022.

\bibitem{Saharia2022PAMI}
C.~Saharia, J.~Ho, W.~Chan, et al., \emph{Image super-resolution via iterative refinement}, IEEE transactions on pattern analysis and machine intelligence, 2022, 45(4): 4713-4726.

\bibitem{Zhang2023CVPR}
L.Zhang, A.~Rao, M.~Agrawala, \emph{Adding conditional control to text-to-image diffusion models}, in Proceedings of the IEEE/CVF International Conference on Computer Vision, pp.3836-3847, 2023.

\bibitem{Ho2022}
J.~Ho, C.~Saharia, W.~Chan, D.J.~Fleet, M.~Norouzi, T.~Salimans, \emph{Cascaded diffusion models for high fidelity image generation}, Journal of Machine Learning Research, 23(47): 1-33, 2022.

\bibitem{Kazerouni}
A.~Kazerouni, E.K.~Aghdam, M.~Heidari, R.~Azad, M.~Fayyaz, I.~Hacihaliloglu, D.~Merhof, \emph{Diffusion models in medical imaging: A comprehensive survey}, Medical Image Analysis, 88: 102846, 2023.

\bibitem{Hein2024}
D.~Hein, A.~Bozorgpour, D.~Merhof, G.~Wang, \emph{Physics-Inspired
Generative Models in Medical Imaging: A Review}, arXiv preprint
arXiv:2407.10856, 2024




\bibitem{Du2024}
W.~Du, H.H.~Cui, L.C.~He, et al. \emph{Structure-aware diffusion for low-dose CT imaging}, Physics in Medicine \& Biology, 2024, 69(15): 155008.

\bibitem{GeMICCAI2023}
R.~Ge, Y.~He, C.~Xia, Y.~Chen, D.~Zhang, G.~Wang, \emph{JCCS-PFGM: A novel circle-supervision based poisson flow generative model for multiphase CECT progressive low-dose reconstruction with joint condition}, in International Conference on Medical Image Computing and Computer-Assisted Intervention, pp. 409-418, 2023. 

\bibitem{Hein2}
D.~Hein, S.~Holmin, T.~Szczykutowicz, J.S.~Maltz, M.~Danielsson, G.~Wang, M.~Persson, \emph{PPFM: Image denoising in photon-counting CT using single-step posterior sampling Poisson flow generative models}, IEEE Transactions on Radiation and Plasma Medical Sciences, 2024.


\bibitem{Restrepo}
C. M.~Restrepo-Galeano, G. R.~Arce, \emph{Super-Resolution in Low Dose X-Ray CT Via Focal Spot Mitigation with Generative Diffusion Networks}, IEEE Transactions on Computational Imaging, 2024.

\bibitem{Karageorgos}
G.M.~Karageorgos, J.~Zhang, N.~Peters, W.~Xia, C.~Niu, H.~Paganetti, G.~Wang, B.De Man, \emph{A denoising diffusion probabilistic model for metal artifact reduction in CT}, IEEE Transactions on Medical Imaging, 2024.

\bibitem{Guan}
B.~Guan, , C.~Yang, L.~Zhang, S.~Niu, M.~Zhang, Y.~Wang, W.~Wu, Q.~Liu, \emph{Generative modeling in sinogram domain for sparse-view CT reconstruction}, IEEE Transactions on Radiation and Plasma Medical Sciences, 2023. 

\bibitem{Liu_ICCV}
J.~Liu, R.~Anirudh, J.J.~Thiagarajan, S.~He, K.A.~Mohan, U.S.~Kamilov, H.~Kim, \emph{DOLCE: A model-based probabilistic diffusion framework for limited-angle ct reconstruction}, In Proceedings of the IEEE/CVF International Conference on Computer Vision, pp. 10498-10508, 2023.


\bibitem{Vincent2011}
P.~Vincent, \emph{A connection between score matching and denoising autoencoders}, Neural computation, 23(7): 1661-1674, 2011.

\bibitem{Scan_protocol}
L.~Yu, M.~Shiung, D.~Jondal, C. H.~McCollough, \emph{Development and validation of a practical lower-dose-simulation tool for optimizing computed tomography scan protocols}, Journal of computer assisted tomography, 2012, 36(4): 477-487.

\bibitem{VinDr}
B. T. ~Dao, T. V. ~Nguyen, H. H.~Pham, H. Q. ~Nguyen, \emph{Phase recognition in contrast‐enhanced CT scans based on deep learning and random sampling}, Medical Physics, 2022, 49(7), 4518-4528.



\end{thebibliography}
\end{document}